\documentstyle[prd,aps,epsf,12pt]{revtex}
\begin{document} 

\tighten
\draft
\preprint{DAMTP96-112} 

\title{Geometrisation of Statistical Mechanics} 
 
\author{
Dorje C. Brody$^{*}$ 
and 
Lane P. Hughston$^{\dagger}$  
} 
\address{$*$Department of Applied Mathematics and 
Theoretical Physics,  \\ University of Cambridge,  
Silver Street, Cambridge CB3 9EW U.K.} 
\address{$\dagger$ Merrill Lynch International, 
25 Ropemaker Street, London EC2Y 9LY U.K. \\ 
and King's College London, The Strand, London 
WC2R 2LS, U.K.}

\date{\today} 

\maketitle 

\begin{abstract} 
{\bf Abstract.} Classical and quantum statistical mechanics are 
cast here in the language of projective geometry to provide a 
unified geometrical framework for statistical 
physics. After reviewing the Hilbert space formulation of 
classical statistical thermodynamics, we introduce projective 
geometry as a basis for analysing probabilistic aspects of 
statistical physics. In particular, the specification of a 
canonical polarity on $RP^{n}$ induces a Riemannian metric 
on the state space of statistical mechanics. In the case of the 
canonical ensemble, we show that equilibrium thermal states are 
determined by the Hamiltonian gradient flow with respect to 
this metric. This flow is concisely characterised by the fact 
that it induces a projective automorphism on the state manifold. 
The measurement problem for thermal systems is studied by the 
introduction of the concept of a random state. The general 
methodology is then extended to include the quantum mechanical 
dynamics of equilibrium thermal states. In this case the relevant 
state space is complex projective space, here regarded as a real 
manifold endowed with the natural Fubini-Study metric. A 
distinguishing feature of quantum thermal dynamics is the inherent 
multiplicity of thermal trajectories in the state space, associated 
with the nonuniqueness of the infinite temperature state. We are 
then led to formulate a geometric characterisation of the 
standard KMS-relation often considered in the context of 
$C^{*}$-algebras. The example of a quantum spin one-half particle 
in heat bath is studied in detail. \par 
\end{abstract} 

\pacs{{\it Keywords: Hilbert space geometry, Projective geometry, 
Equilibrium statistical mechanics, Quantum dynamics} } 


\section{Introduction} 

One of the most fascinating advances in the application of modern 
differential geometry is its use in statistical physics, including 
quantum and statistical mechanics. The purpose of this paper is to 
develop a unified geometrical framework that allows for a natural 
characterisation of both of these aspects of statistical physics. 
\par 

In quantum mechanics, one typically works with square-integrable 
wave functions, i.e., elements of a complex Hilbert space 
${\cal H}$. This space possesses natural geometrical structures 
induced by its norm. However, in order to seek a compelling 
axiomatic formulation of quantum mechanics, it may be reasonable 
to work with a space of more direct physical relevance 
\cite{h-k,lands}. This is not the Hilbert space ${\cal H}$ itself, 
but rather the manifold $\Sigma$ of ``instantaneous pure states'' 
\cite{kibble}, which has the structure of a complex projective 
space $CP^{n}$, possibly infinite dimensional, enriched with a 
Hermitian correspondence, i.e., a complex conjugation operation 
that maps points to hyperplanes in $CP^{n}$, and vice-versa. 
Equivalently, we think of $CP^{n}$ as being endowed with a natural 
Riemannian metric, the Fubini-Study metric. \par 

The space $\Sigma$ is, in fact, the quantum analogue of the 
classical phase space of mechanical systems. Hence, one can 
interpret the Schr\"odinger equation as Hamilton's equations on 
$CP^{n}$, and the equation of motion for a general density matrix 
can be identified with the Liouville equation \cite{gibbons}. The 
advantage of working with the manifold $\Sigma$, rather than the 
Hilbert space of state vectors, above all, is that it can readily 
accommodate generalisations of quantum mechanics \cite{lph1}, 
including nonlinear relativistic models. Furthermore, the 
structure of $\Sigma$ allows for a natural probabilistic 
interpretation even if the standard linear quantum theory is 
modified. \par 

As we discuss elsewhere \cite{dblh0}, the statistical 
aspects of quantum measurement can be greatly clarified if we shift 
our view slightly, and regard the Hilbert space 
${\cal H}$ of quantum mechanics not as a complex Hilbert space, 
but rather a real Hilbert space endowed with a real metric 
and a compatible complex structure. This would appear 
to be simply a change in formalism while keeping the same 
underlying physical structure. Indeed this is so, but once quantum 
theory is formulated this way its relation to other aspects of 
statistical physics becomes much more apparent. \par 

Statistical mechanics, in particular, can also be formulated 
concisely \cite{dblh1} in terms of the geometry of 
a real Hilbert space ${\cal H}$. This can be seen by taking the 
square root of the Gibbs density function, which maps the space of 
probability distributions to vectors in a convex cone ${\cal H}_{+}$ 
in ${\cal H}$. In this way, the various probabilistic and statistical 
operations of statistical mechanics can be given a transparent 
geometric meaning in ${\cal H}$\cite{dblh2,gibbons97}. \par 

However, it can be argued that even at the classical level 
of statistical mechanics the `true' state space is obtained by 
identifying all the pure states along the given ray through the 
origin of ${\cal H}$. In this case, the space obtained is 
essentially the real projective space $RP^{n}$. This is the view 
we take here, and we shall study properties of thermal 
states that become apparent only when the theory is developed in 
a fully geometric context. \par 

The present paper is organised as follows. In Section 2, we 
review the basics of the Hilbert space formulation of statistical 
mechanics. Since this formulation is perhaps not very widely 
appreciated, we can regard this section as an extended 
introduction which then paves the way to the approach in terms 
of projective geometry presented later. We begin with a brief 
review of statistical geometry, including the theory of the 
Fisher-Rao metric on the parameter space of a family of probability 
distributions. The Gibbs distribution 
when viewed in this way can be seen as a curve in Hilbert space, 
parameterised by the inverse temperature, and there is a striking 
formal resemblance to the Schr\"odinger equation, even though here 
we are working at a strictly classical level. \par 

A measurement theory for thermal states is developed by analogy 
with the standard density matrix theory used in quantum mechanics. 
We are then led to a set of uncertainty relations for the 
measurements of thermodynamic conjugate variables such as energy 
and inverse temperature. We also introduce an alternative approach 
to the measurement theory that is not based upon the density matrix 
description. Our approach, based on the introduction of random 
states, extends naturally also to quantum mechanics, where it can 
be seen to be more appealing in a probabilistic context than 
standard treatments, and indeed 
reduces to the conventional density matrix approach in special 
cases. \par 

In Section 3, we introduce a projective geometric framework for 
the probabilistic operations involved in the representation of the 
canonical thermal state associated with the standard Gibbs measure. 
Thermal states are shown to lie on a trajectory in the real 
projective space $RP^{n}$, which is endowed with the natural 
`spherical' metric. In this connection we find it convenient 
to develop a number of useful differential geometric results 
characterising projective transformations on the state space 
$RP^{n}$. We find that the equilibrium thermal trajectories, which 
are shown to be given by a Hamiltonian gradient flow, generate 
projective automorphisms of the state manifold. \par 

In Section 4, we then synthesise the approaches outlined in Sections 
2 and 3, and consider the inter-relationship of the classical 
thermal state space $RP^{n}$ and the quantum phase space $CP^{n}$, 
to study the quantum mechanical dynamics of equilibrium thermal 
states. First we examine the quantum state space from the 
viewpoint of complex algebraic geometry, which shows that this 
space is endowed with a natural Riemannian geometry given by the 
Fubini-Study metric, along with a natural symplectic structure. 
For thermal physics it is instructive to look at quantum mechanics 
from an entirely `real' point of view as well, and this approach is 
developed in Section 4.B. \par 

Our formulation is then compared to the standard 
KMS-construction \cite{hhw} for equilibrium states. In 
particular, once we pass to the mixed state description we 
recover the KMS-state. However, our quantum mechanical pure thermal 
state, which does not obey the KMS-condition, can be viewed as a 
more fundamental construction. In Section 4.D we develop a theory 
of the quantum mechanical microcanonical ensemble, formulated 
entirely in terms of the quantum phase space geometry. This is set 
up in such a way as to admit generalisations to nonlinear 
quantum theories. Finally, we study more explicitly the case of 
a quantum mechanical spin one-half particle in heat bath. \par 

\section{Statistical States in Hilbert Space} 

\subsection{Hilbert space geometry} 

Let us begin by demonstrating how classical statistical mechanics 
can be formulated in an appealing way by the use of a geometrical 
formalism appropriate for Hilbert space. Consider a 
real Hilbert space ${\cal H}$ equipped with an inner 
product $g_{ab}$. A probability density function $p(x)$ can be 
mapped into ${\cal H}$ by taking the square-root 
$\psi(x) = (p(x))^{1/2}$, which is denoted by a vector $\psi^{a}$ in 
${\cal H}$. The normalisation condition $\int (\psi(x))^{2}dx=1$ is 
written $g_{ab}\psi^{a}\psi^{b}=1$, 
indicating that $\psi^{a}$ lies on the unit sphere ${\cal S}$ in 
${\cal H}$. Since a probability density function is nonnegative, the 
image of the map $f: p(x)\rightarrow\psi(x)$ is the 
intersection ${\cal S}_{+}={\cal S}\cap{\cal H}_{+}$ of 
${\cal S}$ with the convex cone 
${\cal H}_{+}$ formed by the totality of quadratically integrable 
nonnegative functions. If we consider the space of all 
probability distributions as a metric space relative to 
the Hellinger distance \cite{xia}, then $f$ is an isometric 
embedding in ${\cal H}$. We call $\psi^{a}$ the state vector of the 
corresponding probability density $p(x)$. \par 

A typical random variable is represented on ${\cal H}$ by a symmetric 
tensor $X_{ab}$, whose expectation in a normalised state 
$\psi^{a}$ is given by 
\begin{equation} 
E_{\psi}[X]\ =\ X_{ab}\psi^{a}\psi^{b}\ . 
\end{equation} 
Similarly, the expectation of its square is 
$X_{ac}X^{c}_{b}\psi^{a}\psi^{b}$. The variance of $X_{ab}$ in the 
state $\psi^{a}$ is therefore ${\rm Var}_{\psi}[X] = {\tilde X}_{ac}
{\tilde X}^{c}_{b} \psi^{a}\psi^{b}$, where ${\tilde X}_{ab} = 
X_{ab} - g_{ab}E_{\psi}[X]$ represents the deviation of $X_{ab}$ from 
its mean in the state $\psi^{a}$. \par 

We consider now the unit sphere $\cal{S}$ in $\cal{H}$, and 
within this sphere a submanifold $\cal{M}$ given parametrically 
by $\psi^{a}(\theta)$, where $\theta^{i}\ (i=1, \cdots,r)$ are 
local parameters. In particular, later on we have in mind the case 
where the parameter space spanned by $\theta^{i}$ represents the 
space of coupling constants in statistical mechanics associated 
with the given physical system. In the case of the canonical Gibbs 
measure there is a single such parameter, corresponding to the inverse 
temperature variable $\beta=1/k_{B}T$. We write $\partial_{i}$ for 
$\partial/\partial\theta^{i}$. Then, in local coordinates, there is 
a natural Riemannian metric ${\cal G}_{ij}$ on the parameter space 
$\cal{M}$, induced by $g_{ab}$, given by 
${\cal G}_{ij} = g_{ab}\partial_{i}\psi^{a} \partial_{j}\psi^{b}$. 
This can be seen as follows. First, note that the squared distance 
between the endpoints of two vectors $\psi^{a}$ and $\eta^{a}$ in 
$\cal{H}$ is $g_{ab} (\psi^{a}-\eta^{a})(\psi^{b}-\eta^{b})$. If both 
endpoints lie on $\cal{M}$, and $\eta^{a}$ is obtained by 
infinitesimally displacing $\psi^{a}$ in $\cal{M}$, i.e., 
$\eta^{a}=\psi^{a}+\partial_{i}\psi^{a}d\theta^{i}$, then the 
separation $ds$ between the two endpoints on $\cal{M}$ is 
$ds^{2} = {\cal G}_{ij}d\theta^{i}d\theta^{j}$, 
where ${\cal G}_{ij}$ is given as above. \par 

The metric ${\cal G}_{ij}$ is, up to 
a conventional, irrelevant factor of four, the so-called 
Fisher-Rao metric on the space of the given family of distributions. 
The Fisher-Rao metric us usually defined in terms of a rather 
complicated expression involving the covariance matrix of the 
gradient of the log-likelihood function; but here we have a 
simple, transparent geometrical construction. 
The Fisher-Rao metric is important since it provides a 
geometrical basis for the key links between the statistical and 
physical aspects of the systems under consideration. \par 

\subsection{Thermal trajectories} 

Now suppose we consider the canonical ensemble of classical 
statistical mechanics, in the case for which the 
system is characterised by a configuration space and an assignment 
of an energy value for each configuration. The parametrised 
family of probability distributions then takes the form of 
the Gibbs measure 
\begin{equation} 
p(H,\beta)\ =\ q(x) \exp\left[- \beta H(x) - W(\beta) 
\right] \ , \label{eq:gib} 
\end{equation} 
where the variable $x$ ranges over the configuration space, 
$H(x)$ represents the energy, $W(\beta)$ is a normalisation 
factor, and $q(x)$ determines the distribution at $\beta=0$, 
where $\beta$ is the inverse temperature parameter. \par 

We 
now formulate a Hilbert space characterisation of this distribution. 
Taking the square-root of $p(H,\beta)$, we find that the state 
vector $\psi^{a}(\beta)$ in ${\cal H}$ corresponding to the Gibbs 
distribution (\ref{eq:gib}) satisfies the differential equation 
\begin{equation} 
\frac{\partial\psi^{a}}{\partial\beta}\ =\ - 
\frac{1}{2}{\tilde H}^{a}_{b} \psi^{b}\ , \label{eq:thermo} 
\end{equation} 
where ${\tilde H}_{ab}=H_{ab}-g_{ab}E_{\psi}[H]$. Here the operator 
$H_{ab}$ in the Hilbert space ${\cal H}$ corresponds to the 
specified Hamiltonian function $H(x)$ appearing in (\ref{eq:gib}). 
The solution of this equation can be represented as follows: 
\begin{equation} 
\psi^{a}(\beta)\ =\ \exp\left[ -\frac{1}{2}(\beta 
H^{a}_{b} + {\tilde W}(\beta)\delta^{a}_{b}) 
\right] q^{b}\ , \label{eq:psi1} 
\end{equation} 
where ${\tilde W}(\beta) = W(\beta)-W(0)$ and $q^{a} = 
\psi^{a}(0)$ is the prescribed distribution at $\beta=0$. \par 

Since $\psi^{a}(\beta)$ respects the normalisation 
$g_{ab}\psi^{a}\psi^{b}=1$, for each value of the temperature $\beta$ 
we find a point on ${\cal M}$ in ${\cal S}_{+}$. To be more specific, 
the thermal system can be described as follows. Consider a unit 
sphere ${\cal S}$ in ${\cal H}$, whose axes label the configurations 
of the system, each of which has a definite energy. We let 
$u^{a}_{k}$ denote an 
orthonormal basis in ${\cal H}$. Here, the index $k$ 
labels all the points in the phase space of the given statistical 
system. In other words, for each point in phase space we have a 
corresponding basis vector $u^{a}_{k}$ in ${\cal H}$ for some 
value of $k$. With this choice of basis, a classical thermal state 
$\psi^{a}(\beta)$ can be expressed as a superposition 
\begin{equation} 
\psi^{a}(\beta)\ =\ e^{-\frac{1}{2}W(\beta)} 
\sum_{k} e^{-\frac{1}{2}\beta E_{k}} u^{a}_{k}\ , 
\label{eq:psi} 
\end{equation} 
where $E_{k}$ is the energy for $k$-th configuration, and thus 
$\exp[W(\beta)]=\sum_{k}\exp(-\beta E_{k})$ is the partition 
function. We note that the states $u_{k}^{a}$ are, in fact, the 
energy eigenstates of the system, with eigenvalues $E_{k}$. That 
is to say, $H^{a}_{b}u^{b}_{k} = E_{k}u^{a}_{k}$. 
The index $k$ in these formulae is formal in the sense that the 
summation may, if appropriate, be replaced by an integration. By 
comparing equations 
(\ref{eq:psi1}) and (\ref{eq:psi}), we find that the initial 
($\beta = 0$) thermal state $q^{a}$ is 
\begin{equation} 
q^{a}\ =\ e^{-\frac{1}{2}W(0)} \sum_{k} u^{a}_{k}\ , 
\end{equation} 
which corresponds to the centre point in ${\cal S}_{+}$. This 
relation reflects the fact that all configurations are equally 
probable likely at infinite temperature. \par 

Viewed as a function of $\beta$, the 
state trajectory $\psi^{a}(\beta)$ thus commences at the centre 
point $q^{a}$, and follows a curve on ${\cal S}$ generated by the 
Hamiltonian $H_{ab}$ according to (\ref{eq:thermo}). It is 
interesting to note that the curvature of this trajectory, given by 
\begin{equation} 
K_{\psi}(\beta)\ =\ \frac{\langle {\tilde H}^{4}\rangle}
{\langle{\tilde H}^{2}\rangle^{2}} 
- \frac{\langle{\tilde H}^{3}\rangle^{2}}
{\langle {\tilde H}^{2}\rangle^{3}} - 1\ , \label{eq:cur} 
\end{equation} 
arises naturally in a physical characterisation of the accuracy 
bounds for temperature measurements. This point 
is pursued further in Section 2.C below, and in \cite{dblh1}. In 
equation (\ref{eq:cur}) the expression 
$\langle {\tilde H}^{n}\rangle$ denotes the $n$-th central 
moment of the observable $H_{ab}$. Here the 
curvature of the curve $\psi^{a}(\beta)$, which is necessarily 
positive, is the square of the `acceleration' vector 
along the state trajectory $\psi^{a}(\beta)$, normalised by the 
square of the velocity vector. \par 

\subsection{Measurement for thermal states} 

Given the thermal trajectory $\psi^{a}(\beta)$ above, we propose, 
in the first instance, to consider measurement and estimation 
by analogy with the von Neumann approach in quantum mechanics. 
According to this scheme the {\it general} state of a 
thermodynamic system is represented by a `density matrix' 
$\rho^{ab}$ which in the present context should be understood to 
be a symmetric, semidefinite matrix with trace unity; that is to 
say, $\rho^{ab}\xi_{a}\xi_{b} \geq 0$ for any covector $\xi_{a}$, 
and $\rho^{ab}g_{ab}=1$. Then, for example, we can write 
\begin{equation} 
E_{\rho}[X]\ =\ X_{ab}\rho^{ab} 
\end{equation} 
for the expectation of a random variable 
$X_{ab}$ in the state $\rho^{ab}$, and 
\begin{equation} 
{\rm Var}_{\rho}[X]\ =\ X_{ab}X^{b}_{c}\rho^{ac} - 
(X_{ab}\rho^{ab})^{2} 
\end{equation} 
for the variance of $X_{ab}$ in that state. 
It should be evident that in the case of a 
pure state, for which $\rho^{ab}$ is of the form $\rho^{ab} = 
\psi^{a}\psi^{b}$ for some state vector $\psi^{a}$, these formulae 
reduce to the expressions considered in Section 2.A. \par 

In particular, let us consider measurements made on a pure 
equilibrium state $\psi^{a}(\beta)$. Such measurements are 
characterised by projecting the prescribed state onto the ray in the 
Hilbert space corresponding to a specified point in the phase space. 
Hence, the probability of observing the $k$-th state, when the 
system is in the pure state $\psi^{a}$,  is given by the 
corresponding Boltzmann weight 
\begin{equation} 
p_{k}\ =\ (g_{ab}\psi^{a}u^{b}_{k})^{2}\ =\ 
e^{-\beta E_{k}-W(\beta)}\ . 
\end{equation} 
In terms of the density matrix description, the state before 
measurement is given by the degenerate pure state matrix $\rho^{ab} 
= \psi^{a}\psi^{b}$, for which the thermal development is 
\begin{equation} 
\frac{d\rho^{ab}}{d\beta}\ =\ - \frac{1}{2}\left( 
{\tilde H}^{a}_{c}\rho^{bc} + {\tilde H}^{b}_{c}\rho^{ac} 
\right) \ , 
\end{equation} 
or equivalently $d\rho/d\beta = - \{ {\tilde H},\rho\}$, where 
$\{ A,B \}$ denotes the symmetric product between the 
operators $A$ and $B$. After a measurement, $\rho^{ab}$ takes 
the form of a mixed state, characterised by a 
nondegenerate diagonal density operator for which the diagonal 
elements are the Boltzmann weights $p_{k}$. In 
this state vector reduction picture, the von Neumann entropy 
$-{\rm Tr}[\rho\ln\rho]$ changes from $0$ to its maximum value 
$S=\beta\langle H\rangle+W_{\beta}$, which can be viewed as the 
quantity of information gained from the observation.  \par 

More generally, suppose we consider the measurement of an 
arbitrary observable $X_{ab}$ in the state $\psi^{a}(\beta)$ in 
the situation when the spectrum of $X_{ab}$ admits a continuous 
component. In 
this case, we consider the spectral measure associated with the 
random variable $X_{ab}$. Then, the probability density for the 
measurement outcome $x$ is given by the expectation 
$p(x,\beta) = \Pi_{ab}(X,x)\psi^{a}\psi^{b}$ of the 
projection operator 
\begin{equation}  
\Pi^{a}_{b}(X,x)\ =\ \frac{1}{\sqrt{2\pi}} 
\int_{-\infty}^{\infty} \exp\left[ i\lambda(X^{a}_{b} - 
x\delta^{a}_{b}) \right] d\lambda  \ . 
\end{equation}   
In other words, we assign a projection-valued measure $\Pi_{X}(x)$ 
on the real line associated with each symmetric operator $X$, so 
that for a given unit vector $\psi^{a}$, the mapping 
$x\in{\bf R}\mapsto E_{\psi}[\Pi_{X}(x)]$ is a probability measure. 
This measure determines the distribution of values obtained when 
the observable $X$ is measured while the system is 
in the state $\psi$. \par 

For a more refined view of the measurement problem we need to 
take into account some ideas from statistical estimation theory. 
Suppose that we want to make a measurement or series of 
measurements to estimate the value of the parameter characterising 
a given thermal equilibrium state. In this situation the observable 
we measure is called an `estimator' for the given parameter. We 
are interested in the case for which the estimator is unbiased in 
the sense that its expectation gives the value of the required 
parameter. To be specific, we consider the case when we estimate 
the value of the temperature. Let 
$B_{ab}$ be an unbiased estimator for $\beta$, so 
that along the trajectory $\psi^{a}(\beta)$ we have: 
\begin{equation} 
\frac{B_{ab}\psi^{a}\psi^{b}}
{g_{cd}\psi^{c}\psi^{d}}\ =\ \beta\ . 
\end{equation} 
As a consequence of this relation and the thermal state 
equation (\ref{eq:thermo}), we observe that the inverse temperature 
estimator $B$ and the Hamiltonian $H$ satisfy the `weak' 
anticommutation 
relation $E_{\psi}[\{ B,{\tilde H}\}] = -1$ along the state 
trajectory $\psi$. In 
statistical terms, this implies that these conjugate variables 
satisfy the covariance relation 
$E_{\psi}[BH]-E_{\psi}[B]E_{\psi}[H]=-1$ along the 
trajectory. \par 

\subsection{Thermodynamic uncertainty relations} 

Equipped with the above definitions, one can easily verify that the 
variance in estimating the inverse temperature parameter $\beta$ can 
be expressed by the geometrical relation 
\begin{equation} 
{\rm Var}_{\psi}[B]\ =\ \frac{1}{4} 
g^{ab}\nabla_{a}\beta\nabla_{b}\beta \label{eq:gr} 
\end{equation} 
on the unit sphere ${\cal S}$, where $\nabla_{a}\beta = 
\partial\beta/\partial\psi^{a}$ is the gradient of the 
temperature estimate $\beta$. The essence of formula (\ref{eq:gr}) 
can be understood as follows. 
First, recall that $\beta$ is the expectation 
of the estimator $B_{ab}$ in the state $\psi^{a}(\beta)$. 
Suppose that the state changes rapidly as $\beta$ changes. Then, the 
variance in estimating $\beta$ is small---indeed, this is given by 
the squared magnitude of the `functional derivative' of $\beta$ with 
respect to the state $\psi^{a}$. On the other hand, if the state 
does not change significantly as $\beta$ changes, then the 
measurement outcome of an observable is less conclusive in 
determining the value of $\beta$. \par 

The squared length of the gradient vector $\nabla_{a}\beta$ can 
be expressed as a sum of squares of orthogonal components. To this 
end, we choose a new set of orthogonal basis vectors 
given by the state $\psi^{a}$ and its higher derivatives. 
If we let $\psi^{a}_{n}$ denote $\psi^{a}$ for $n=0$, and 
for $n>0$ the component of the derivative $\partial^{n}\psi^{a}/
\partial\beta^{n}$ orthogonal to the state $\psi^{a}$ and its 
lower order derivatives, then our orthonormal vectors are 
given by ${\hat \psi}^{a}_{n} = \psi^{a}_{n}
(g_{bc}\psi^{b}_{n}\psi^{c}_{n})^{-1/2}$ for $n=0,1,2,\cdots$. 
With this choice of orthonormal vectors, we find that the variance 
of the estimator $B$ satisfies  
\begin{equation} 
{\rm Var}_{\psi}[B]\ \geq\ \sum_{n} 
\frac{({\tilde B}_{ab}\psi^{a}_{n}\psi^{b})^{2}}
{g_{cd}\psi^{c}_{n}\psi^{d}_{n}} \ , 
\label{eq:vb} 
\end{equation} 
for any range of the index $n$. This follows because 
the squared magnitude of the vector $\frac{1}{2} 
\nabla_{a}\beta = {\tilde B}_{ab} \psi^{b}$ is 
greater than or equal to the sum of the squares of its 
projections onto the basis vectors given by 
${\hat \psi}^{a}_{n}$ for the specified range of $n$. \par 

In particular, for $n=1$ we have $B_{ab}\psi^{a}_{1}\psi^{b} = 
\frac{1}{2}$ on account of the relation $B_{ab}\psi^{a}\psi^{b} = 
\beta$, and $g_{ab}\psi^{a}_{1}\psi^{b}_{1}=\frac{1}{4} 
\Delta H^{2}$, which follows from the thermal equation 
(\ref{eq:thermo}). Therefore, if we write ${\rm Var}_{\psi}[B] = 
\Delta\beta^{2}$, we find for $n=1$ that the inequality (\ref{eq:vb}) 
implies the following thermodynamic uncertainty relation: 
\begin{equation} 
\Delta\beta^{2}\Delta H^{2}\ \geq\ 1 \ , \label{eq:tu} 
\end{equation} 
valid along the trajectory consisting of the thermal equilibrium 
states. The variance $\Delta^{2}\beta$ here is to be understood in 
the sense of estimation theory. That is, although the variable 
$\beta$ does not actually fluctuate, as should be clear from the 
definition of canonical ensemble, there is nonetheless an 
inevitable lower bound for the variance of the measurement, 
given by (\ref{eq:tu}), if we wish to estimate 
the value of the heat bath temperature $\beta$. It is worth 
pointing out that the exposition we have given here is consistent 
with the view put forward by Mandelbrot \cite{man}, who should 
perhaps be credited with first having introduced 
an element of modern statistical reasoning into the long-standing 
debate on the status of temperature fluctuations \cite{man2}. \par 

Note that, although we only considered the variance 
$\langle (B-\langle B\rangle)^{2}\rangle$ here, 
the higher order central moments $\mu_{n} = 
\langle (B-\langle B\rangle)^{n}\rangle$ can also be expressed 
geometrically. This can be seen as follows. First, 
recall that for any observable $F_{ab}$ with 
$E_{\psi}[F] = f$, we have $\nabla_{a}f = 
2{\tilde F}_{ab}\psi^{b}$ on the unit sphere ${\cal S}$. 
Therefore, by letting $F = {\tilde B}^{n}$, we 
construct the higher central moments in terms of the 
cosines of the angles between certain gradient vectors, e.g., 
$4\mu_{3} = g^{ab}\nabla_{a}\mu_{2}\nabla_{b}\beta$,  
$4\mu_{4} = g^{ab}\nabla_{a}\mu_{2}\nabla_{b}\mu_{2}-4\mu_{2}^{2}$, 
and so on. In particular, the even order moments are expressible 
in terms of combinations of the squared lengths of normal vectors 
to the surfaces of constant central moments of lower order. \par 

\subsection{Random states} 

Let us return to the consideration of measurements on thermal 
states, which we now pursue in greater depth. In doing so 
we shall introduce the idea of a `random state', a concept that 
is applicable both in clarifying the measurement problem in 
statistical physics, as well as in providing a useful tool when we 
consider ensembles. It also turns out that the idea of a random 
state is helpful in the analysis of conceptual problems in quantum 
mechanics. Later on when we consider quantum statistical mechanics, 
we shall have more to say on this. \par 

Suppose we consider a pure thermal state $\psi^{a}(\beta)$ for 
some value $\beta$ of the inverse temperature. We know that 
this state is given by 
\begin{equation} 
\psi^{a}(\beta)\ =\ \sum_{k}p_{k}^{1/2}u_{k}^{a}\ , 
\end{equation} 
where $u^{a}_{k}$ is a normalised energy eigenstate with eigenvalue 
$E_{k}$, and $p_{k}$ is the associated Gibbs probability. After 
measurement, it is natural to consider the outcome of the 
measurement to be a {\it random state} ${\bf \Psi}^{a}$. Thus we 
consider ${\bf \Psi}^{a}$ to be a random variable (indicated by 
use of a bold font) such that the probability for taking a 
given eigenstate is 
\begin{equation} 
{\rm Prob}[{\bf \Psi}^{a} = u^{a}_{k}]\ =\ p_{k}\ . 
\end{equation} 
This way of thinking about the outcome of the measurement 
process is to some extent complementary to the density matrix 
approach, though in what follows we shall make it clear what the 
relationship is. \par 

In particular, the expectation of an observable $X_{ab}$ in the 
random state ${\bf \Psi}^{a}$ is given by averaging over the 
random states, that is, 
\begin{equation} 
E_{{\bf \Psi}}[X]\ =\ X_{ab}{\bf \Psi}^{a}{\bf \Psi}^{b}\ . 
\end{equation} 
This relation should be interpreted as the specification of a 
{\it conditional} expectation, i.e., the conditional expectation 
of $X_{ab}$ in the random state ${\bf \Psi}^{a}$. Then the 
associated {\it unconditional} expectation 
$E[X]=E[E_{\bf \Psi}[X]]$ is given by 
\begin{equation} 
E[X]\ =\ X_{ab}E[{\bf \Psi}^{a}{\bf \Psi}^{b}]\ . 
\end{equation} 
However, since ${\rm Prob}[{\bf \Psi}^{a}=u^{a}_{k}]=p_{k}$, it 
should be evident that 
\begin{equation} 
E[{\bf \Psi}^{a}{\bf \Psi}^{b}]\ =\ \rho^{ab}\ , \label{eq:ue} 
\end{equation} 
where the density matrix $\rho^{ab}$ is defined by 
\begin{equation} 
\rho^{ab}\ =\ \sum_{k}p_{k}u^{a}_{k}u^{b}_{k}\ . 
\end{equation} 
Thus, the unconditional expectation of the random 
variable $X_{ab}$ is given, as noted earlier, by $E[X] = 
X_{ab}\rho^{ab}$. It should be observed, however, that here we 
are not emphasising the role of the density matrix 
$\rho^{ab}$ as representing a `state', but rather its role in 
summarising information relating to 
the random state ${\bf \Psi}^{a}$. \par 

The feature that distinguishes the density matrix in this analysis 
is that it is fully sufficient for the characterisation of 
unconditional statistics relating to the observables and states 
under consideration. This point is clearly illustrated when we 
calculate the variance of a random variable $X_{ab}$ in a random 
state ${\bf \Psi}^{a}$. Such a situation arises if we want 
to discuss the uncertainties arising in the measurement of an 
observable $X_{ab}$ for an ensemble. In this case the system we 
have in mind is a large number of identical, independent 
particles, each of which is in a definite energy 
eigenstate, where the distribution of the energy is given 
according to the Gibbs distribution. One might take this as an 
elementary model for a classical gas. Then the distribution of the 
ensemble can be described in terms of a random state ${\bf \Psi}^{a}$. 
Note that here the interpretation is slightly different from what 
we had considered before (the random outcome of a measurement for an 
isolated system), though it will be appreciated that the relation 
of these two distinct interpretations is of considerable interest 
for physics and statistical theory alike. \par 

The conditional variance of the observable $X_{ab}$ in the random 
state ${\bf \Psi^{a}}$ is given by 
\begin{equation} 
{\rm Var}_{\bf \Psi}[X]\ =\ X_{ab}X^{b}_{c}
{\bf \Psi}^{a}{\bf \Psi}^{c} - 
(X_{ab}{\bf \Psi}^{a}{\bf \Psi}^{b})^{2}\ . 
\end{equation} 
The average over the different values of ${\bf \Psi}^{a}$ then 
gives us 
\begin{equation} 
E\left[ {\rm Var}_{\bf \Psi}[X]\right]\ =\ 
X_{ab}X^{b}_{c}\rho^{ac} - X_{ab}X_{cd}\rho^{abcd}\ , 
\end{equation} 
where $\rho^{ab}$ is, as before, the density matrix (\ref{eq:ue}), 
and $\rho^{abcd}$ is a 
certain higher moment of ${\bf \Psi}^{a}$, defined by 
\begin{equation} 
\rho^{abcd}\ =\ E[{\bf \Psi}^{a}{\bf \Psi}^{b}
{\bf \Psi}^{c}{\bf \Psi}^{d}]\ . 
\end{equation} 
\par 

The appearance of this higher order analogue of the density matrix 
may be surprising, though it is indeed a characteristic 
feature of conditional probability. However, the unconditional 
variance of $X$ is not given simply by the expectation 
$E\left[{\rm Var}_{\bf \Psi}[X]\right]$, but rather (see, e.g., 
\cite{ross}) by the {\it conditional variance formula} 
\begin{equation} 
{\rm Var}[X]\ =\ E\left[{\rm Var}_{\bf \Psi}[X]\right] + 
{\rm Var}[E_{\bf \Psi}[X]]\ . \label{eq:uvar} 
\end{equation} 
For the second term we have 
\begin{eqnarray} 
{\rm Var}[E_{\bf \Psi}[X]]\ &=&\ E[(X_{ab}{\bf \Psi}^{a}
{\bf \Psi}^{b})^{2}] - 
(E[X_{ab}{\bf \Psi}^{a}{\bf \Psi}^{b}])^{2}\nonumber \\ 
&=&\ X_{ab}X_{cd}\rho^{abcd} - (X_{ab}\rho^{ab})^{2}\ , 
\end{eqnarray} 
which also involves the higher moment $\rho^{abcd}$. The 
terms in (\ref{eq:uvar}) involving $\rho^{abcd}$ then cancel, 
and we are left with ${\rm Var}[X] = X_{ab}X^{b}_{c}\rho^{ac} 
- (X_{ab}\rho^{ab})^{2}$ for the unconditional variance, which, as 
indicated earlier, only involves the density matrix. It follows 
that the random state approach does indeed reproduce the earlier 
density matrix formulation of our theory, though the role of the 
density matrix is somewhat diminished. In other words, whenever 
conditioning is involved, it is the set of totally symmetric 
tensors 
\begin{equation} 
\rho^{ab\cdots cd}\ =\ 
E[{\bf \Psi}^{a}{\bf \Psi}^{b}\cdots{\bf \Psi}^{c}{\bf \Psi}^{d}] 
\end{equation} 
that plays the fundamental role, although it suffices to consider 
the standard density matrix $\rho^{ab}$ when conditioning is 
removed. \par 

All this is worth having in mind later when we turn to quantum 
statistical mechanics, where the considerations we have developed 
here in a classical context reappear in a new light. 
We want to de-emphasise the role of the density 
matrix, not because there is anything wrong {\it per se} with the 
use of the density matrix in an appropriate context, but rather for 
two practical reasons. First of all, when we want to 
consider conditioning, exclusive attention on the density matrix 
hampers our thinking, since, as we have indicated, higher moments of 
the random state vector also have a role to play. Second, when we 
go to consider generalisations of quantum mechanics, such as the 
nonlinear theories of the Kibble-Weinberg type 
\cite{kibble78,weinberg}, or stochastic theories of the type 
considered by Gisin, Percival, and others \cite{gisin,lph}, the density 
matrix is either an ill formulated concept, or plays a diminished 
role. We shall return to this point for further discussion when 
we consider quantum statistical mechanics in Section 4. \par 
 
\section{Statistical Phase Space} 

\subsection{Projective space and probabilities} 

To proceed further it will be useful to develop a formalism for 
the algebraic treatment of real projective geometry, with a view to 
its probabilistic interpretation in the context of classical 
statistical mechanics. 
Let $Z^{a}$ be coordinates for $(n+1)$-dimensional real Hilbert 
space ${\cal H}^{n+1}$. Later, when we consider quantum theory from 
a real point of view we shall double this dimension. In the Hilbert 
space description of 
classical probabilities, the normalisation condition is written 
$g_{ab}Z^{a}Z^{b}=1$. However, this normalisation is physically 
irrelevant since the expectation of an arbitrary operator $F_{ab}$ 
is defined by the ratio 
\begin{equation} 
\langle F\rangle\ =\ \frac{F_{ab}Z^{a}Z^{b}}{g_{cd}Z^{c}Z^{d}}\ . 
\end{equation} 
Therefore, the physical state space is not the Hilbert space 
${\cal H}$, but the space of equivalence classes 
obtained by identifying the points $\{Z^{a}\}$ and $\{\lambda 
Z^{a}\}$ for all $\lambda \in {\bf R}-\{0\}$. In this way, we 
`gauge away' the irrelevant degree of freedom. The resulting space is 
the real projective $n$-space $RP^{n}$, the space of 
rays through the origin in ${\cal H}^{n+1}$. Thus, two points $X^{a}$ 
and $Y^{a}$ in ${\cal H}^{n+1}$ are equivalent in $RP^{n}$ if they 
are proportional, i.e., $X^{[a}Y^{b]}=0$. \par 

The coordinates $Z^{a}$ (excluding $Z^{a}=0$) can be used as 
homogeneous coordinates for points of $RP^{n}$. Clearly $Z^{a}$ 
and $\lambda Z^{a}$ represent the same point in $RP^{n}$. In 
practice one treats the homogeneous coordinates as though they define 
points of ${\cal H}^{n+1}$, with the stipulation that the allowable 
operations of projective geometry are those which transform 
homogeneously under the rescaling $Z^{a}\rightarrow\lambda Z^{a}$. \par 

A {\it prime}, or $(n-1)$-plane in $RP^{n}$ consists of a set of 
points $Z^{a}$ which satisfy a linear equation $P_{a}Z^{a}=0$, 
where we call $P_{a}$ the homogeneous coordinates of the prime. 
Clearly, $P_{a}$ and $\lambda P_{a}$ determine the same prime. 
Therefore, a prime in $RP^{n}$ is an $RP^{n-1}$, and the set of all 
primes in $RP^{n}$ is itself an $RP^{n}$ (the `dual' projective 
space). In particular, the metric $g_{ab}$ on ${\cal H}^{n+1}$ can 
be interpreted in the projective space as giving rise to a 
nonsingular {\it polarity}, that is, an invertible map from points 
to hyperplanes in $RP^{n}$ of codimension one. This map is given by 
$P^{a}\rightarrow P_{a}:= g_{ab}P^{b}$. See reference \cite{hh} for 
further discussion of some of the geometric operations employed 
here. \par 

If a point 
$P^{a}$ in $RP^{n}$ corresponds to a probability state, then its 
negation $\neg P^{a}$ is the hyperplane $P_{a}Z^{a}=0$. To be 
more precise, we take $P^{a}$ as describing the probability state 
for a set of events. Then, all the probability states 
corresponding to the complementary events lie in the prime 
$P_{a}Z^{a}=0$. Thus, the points $Z^{a}$ on this plane are 
precisely the states that are orthogonal to 
the original state $P^{a}$. The intuition behind this is as 
follows. Two states $P^{a}$ and $Q^{a}$ are orthogonal if and only 
if any event which in the state $P^{a}$ (resp. $Q^{a}$) has a 
positive probability is assigned zero probability by the state 
$Q^{a}$ (resp. $P^{a}$). This is the sense in which orthogonal 
states are `complementary'. For any state $P^{a}$, the plane
consisting of all points $Z^{a}$ such that $P_{a}Z^{a}=0$ is the 
set of states complementary to $P^{a}$ in this sense. \par 

Distinct states $X^{a}$ and $Y^{a}$ are 
joined by a real projective line represented by the skew tensor 
$L^{ab}=X^{[a}Y^{b]}$. The points on this line are the various real 
superpositions of the original two states. The intersection of the 
line $L^{ab}$ with the plane $R_{a}$ is given by $S^{a} = 
L^{ab}R_{b}$. Clearly $S^{a}$ lies on the plane $R_{a}$, since 
$R_{a}S^{a}=0$ on account of the antisymmetry of $L^{ab}$. \par 

The hyperplanes that are the negations (polar planes) of two points 
$P^{a}$ and $Q^{a}$ intersect the joining line $L^{ab}=P^{[a}Q^{b]}$ 
at a pair of points ${\tilde P}^{a}$ and ${\tilde Q}^{a}$ 
respectively. That is, if $P_{a}=g_{ab}P^{b}$ is the coordinate of a 
plane and $L^{ab}$ represents a line in $RP^{n}$, then the point of 
intersection is given by ${\tilde P}^{a} = L^{ab}P_{b}$, and 
similarly ${\tilde Q}^{a}=L^{ab}Q_{b}$. 
The projective cross ratio between 
these four points $P^{a}$, $Q^{a}$, ${\tilde P}^{a} = P^{a} 
(Q^{b}P_{b})-Q^{a}(P^{b}P_{b})$ and ${\tilde Q}^{a} = P^{a} 
(Q^{b}Q_{b})-Q^{a}(P^{b}Q_{b})$, given by $P^{a}{\tilde Q}_{a} 
Q^{b}{\tilde P}_{b}/P^{c}{\tilde P}_{c}Q^{d}{\tilde Q}_{d}$, 
reduces, after some algebra, to the following simple expression: 
\begin{equation} 
\kappa\ =\ \frac{(P^{a}Q_{a})^{2}} 
{P^{b}P_{b}Q^{c}Q_{c}}\ , \label{eq:tcr} 
\end{equation} 
which can be interpreted as the transition probability between 
$P^{a}$ and $Q^{a}$. It is interesting to note that this formula has 
an analogue in quantum mechanics \cite{lph1}. \par 

The projective line $L^{ab}$ can also be viewed as a circle, with 
${\tilde P}^{a}$ and ${\tilde Q}^{a}$ antipodal to the points 
$P^{a}$ and $Q^{a}$. In that case, the cross ratio $\kappa$ is 
$\frac{1}{2}(1 + \cos\theta)=\cos^{2}(\theta/2)$ where $\theta$ 
defines the angular distance between $P^{a}$ and $Q^{a}$, in the 
geometry of $RP^{n}$. We note that $\theta$ is, in fact, {\it twice} 
the angle made between the corresponding Hilbert space vectors, so 
orthogonal states are maximally distant from one another. \par 

Now suppose we let the two states $P^{a}$ and $Q^{a}$ approach one 
another. In the limit the resulting formula for their second-order 
infinitesimal separation determines the natural line element on real 
projective space. This can be obtained by setting $P^{a}=Z^{a}$ and 
$Q^{a}=Z^{a} + dZ^{a}$ in (\ref{eq:tcr}), while replacing $\theta$ 
with the small angle $ds$ in the expression 
$\frac{1}{2}(1+\cos\theta)$, retaining terms of the second order in 
$ds$. Explicitly, we obtain 
\begin{equation} 
ds^{2}\ =\ 4\left[ \frac{dZ^{a}dZ_{a}}{Z^{b}Z_{b}} - 
\frac{(Z^{a}dZ_{a})^{2}}{(Z^{b}Z_{b})^{2}}\right] \ . \label{eq:rpm2} 
\end{equation} 
Note that this metric \cite{kn} is related to the metric on the 
sphere ${\cal S}^{n}$ in ${\cal H}^{n+1}$, except that in the case of 
the sphere one does not identify opposite points. \par 

We now consider the case where the real projective space is 
the state space of classical statistical mechanics. If we write 
$\psi^{a}(\beta)$ for the trajectory of thermal 
state vectors, as discussed in Section 2, then we can regard 
$\psi^{a}(\beta)$, for each value of $\beta$, as representing 
homogeneous coordinates for points in the state space $RP^{n}$. 
Since $\psi^{a}(\beta)$ satisfies (\ref{eq:thermo}) it follows 
that the line element along the curve $\psi^{a}(\beta)$ is given by 
$ds^{2} = \langle{\tilde H}^{2}\rangle d\beta^{2}$, 
which can be identified with the Fisher-Rao metric induced on the 
thermal trajectory by virtue of its embedding in $RP^{n}$. This 
follows by insertion of the thermal equation (\ref{eq:thermo}) 
into expression (\ref{eq:rpm2}) for the natural spherical 
metric on $RP^{n}$. \par 

\subsection{The projective thermal equations} 

Let us write $H_{ab}$ for the symmetric Hamiltonian operator, and 
$\psi^{a}(\beta)$ for the one-parameter family of thermal states. 
Then in our notation the thermal equation is $d\psi^{b} = 
-\frac{1}{2}{\tilde H}^{b}_{c}\psi^{c}d\beta$. However, we are 
concerned with this equation only inasmuch as it supplies 
information about the evolution of the state of the system, 
i.e., its motion in $RP^{n}$. We are interested therefore 
primarily in the {\it projective} thermal equation, given by 
\begin{equation} 
\psi^{[a}d\psi^{b]}\ =\ -\frac{1}{2} \psi^{[a}{\tilde H}^{b]}_{c} 
\psi^{c}d\beta \ , \label{eq:p-ts} 
\end{equation} 
obtained by skew-symmetrising the thermal equation (\ref{eq:thermo}) 
with $\psi^{a}$. Equation (\ref{eq:p-ts}) defines the equilibrium 
thermal trajectory of a statistical mechanical system in the proper 
state space. \par 

The thermal equation generates a Hamiltonian 
gradient flow on the state manifold. This can be seen as follows. 
First, recall that a physical observable $F$ is associated with a 
symmetric operator $F_{ab}$, and the set of such observables form 
a vector space of dimension $\frac{1}{2}(n+1)(n+2)$. Such 
observables are determined by their diagonal matrix elements, which 
are real valued functions on $RP^{n}$ of the form 
$F(\psi^{a}) = F_{ab}\psi^{a}\psi^{b}/\psi^{c}\psi_{c}$. 
In particular, we are interested in the Hamiltonian function 
$H(\psi^{a})$. Then, by a direct 
substitution, we find that the vector field $H^{a} = 
g^{ab}\partial H/\partial \psi^{b}$ associated with the Hamiltonian 
function $H$ takes the form $H^{a}=2{\tilde H}^{a}_{b}\psi^{b}/ 
\psi^{c}\psi_{c}$. Therefore, 
we can write the differential equation for the thermal 
state trajectory in ${\cal H}^{n+1}$ in the form 
\begin{equation} 
\frac{d\psi^{a}}{d\beta}\ =\ -\frac{1}{4}g^{ab}\nabla_{b}H \ . 
\end{equation} 
By projecting this down to $RP^{n}$, we obtain 
\begin{equation} 
\psi^{[a}d\psi^{b]}\ =\ -\frac{1}{4}\psi^{[a}\nabla^{b]}Hd\beta\ , 
\end{equation} 
where $\nabla^{a}H=g^{ab}\nabla_{b}H$. From this we can then 
calculate the line element to obtain  
\begin{eqnarray} 
ds^{2}\ &=&\ 8\psi^{[a}d\psi^{b]}\psi_{[a}d\psi_{b]}/
(\psi^{c}\psi_{c})^{2} \nonumber \\ 
&=&\ \frac{1}{2}\psi^{[a}\nabla^{b]}H \psi_{[a}\nabla_{b]}H 
d\beta^{2}\nonumber \\ 
&=&\ \frac{1}{4}\nabla^{b}H\nabla_{b}H\psi^{a}\psi_{a} 
d\beta^{2}\nonumber \\ 
&=&\ \langle{\tilde H}^{2}\rangle d\beta^{2}\ , 
\end{eqnarray} 
which establishes the result we noted earlier. \par  

The critical points of the Hamiltonian function are the fixed points 
in the state space associated with the gradient vector field 
$g^{ab}\nabla_{b}H$. In the case of Hamilton's equations the fixed 
points are called stationary states. In the Hilbert 
space ${\cal H}^{n+1}$ these are the points corresponding 
to the energy eigenstates $u^{a}_{k}$ given by $H^{a}_{b}u^{b}_{k} = 
E_{k}u^{a}_{k}$ (in the general situation with distinct energy 
eigenvalues $E_{k}$, $k=0,1,\cdots,n$). 
Therefore, in the projective space $RP^{n}$ we have a set of 
fixed points corresponding to the states $u^{a}_{k}$, and the 
thermal states are obtained by superposing these points with an 
appropriate set of coefficients given by the Boltzmann weights. 
Since these coefficients are nonzero at finite temperature, it should 
be clear that the thermal trajectories do not intersect any of 
these fixed points. \par 

In particular, the infinite temperature $(\beta = 0)$ thermal state 
$\psi^{a}(0)$ is located at the centre point of ${\cal S}_{+}$ in 
${\cal H}^{n+1}$. The distances from $\psi^{a}(0)$ to the various 
energy eigenstates are all equal. This implies that in $RP^{n}$ the 
cross ratios between the fixed points and the state $\psi^{a}(0)$ 
are equal. Therefore, we can single out the point $\psi^{a}(0)$ as 
an initial point, and form a geodesic hypersphere in $RP^{n}$. All 
the fixed points of the Hamiltonian vector field $H^{a}$ lie on this 
sphere. Since the cross ratios between the fixed points are also 
equal (i.e., maximal), these fixed points form a regular simplex on 
the sphere. The thermal trajectory thus commences at $Z^{a}(0)$, and 
asymptotically approaches a fixed point associated with the lowest 
energy eigenvalue $E_{0}$, as $\beta \rightarrow \infty$. \par 

If we take the orthogonal prime of the thermal state for any finite 
$\beta$, i.e., $\neg \psi^{a}(\beta)$, then the resulting hyperplane 
clearly does not contain any of the fixed points. On the other hand, 
if we take the orthogonal prime of any one of the fixed points 
$u^{a}_{k}$, then the resulting hyperplane includes a sphere of 
codimension one where all the other fixed points lie. This sphere is 
given by the intersection of the original hypersphere surrounding 
$\psi^{a}(0)$ with the orthogonal prime of the given excluded 
fixed point. 
There is a unique prime containing $n$ general points in 
$RP^{n}$. It is worth noting, therefore, that if we choose 
$n$ general points given by $u^{a}_{j}$ ($j=0,1,\cdots,n$;$j\neq k$), 
then there is a unique solution, up to proportionality, of the $n$ 
linear equations $X_{a}u^{a}_{0}=0$, $\cdots$, $X_{a}u^{a}_{n}=0$. 
The solution is then given by $X^{a}=u^{a}_{k}$, where 
$u^{a}_{k} = \epsilon^{a}_{bc\cdots d}u^{b}_{0}u^{c}_{1}\cdots 
u^{d}_{n}$. Here, 
$\epsilon_{ab\cdots c}=\epsilon_{[ab\cdots c]}$, with $n+1$ 
indices, is the totally skew tensor determined up to 
proportionality. \par 

It would be interesting to explore whether the orientability 
characteristics of $RP^{n}$ lead to any physical consequences. There 
may be a kind of purely classical `spin-statistics' relation in the 
sense that the state space of half-integral spins are associated with 
a topological invariant, while the state space for even spins are 
not. \par 

\subsection{Hamiltonian flows and projective transformations} 

We have seen in Section 3.B how the thermal trajectory of a 
statistical mechanical system is generated by the gradient flow 
associated with the Hamiltonian function. 
In other words, for each point on the state 
manifold we form the expectation of the Hamiltonian in that state. 
This gives us a global function on the state manifold, which we 
call the Hamiltonian function. Next, we take the gradient of this 
function, and raise the index by use of the natural metric to 
obtain a vector field. This vector field is the generator of the 
thermal trajectories. \par 

It is interesting to note that there is a relation between the 
geometry of such vector fields and the global symmetries of the 
state manifold. In particular, we shall 
show below that gradient vector fields generated by 
observables on the state manifold can also be interpreted as the 
generators of {\it projective transformations}. 
A projective transformation on a Riemannian 
manifold is an automorphism that maps geodesics 
onto geodesics. In the case of a real projective space endowed 
with the natural metric, the general such automorphism is 
generated, as we shall demonstrate, by a vector field that is 
expressible in the form of a sum of a Killing vector 
and a gradient flow associated with an observable function. \par 

To pursue this point further we develop some differential 
geometric aspects of the state manifold. We consider $RP^{n}$ 
now to be a differential manifold endowed with the natural 
spherical metric $g_{\bf ab}$. Here, bold upright indices signify 
local tensorial operations in the tangent space of this manifold. 
Thus we write $\nabla_{\bf a}$ for the covariant derivative 
associated with $g_{\bf ab}$, and for an 
arbitrary vector field $V^{\bf a}$ we define the 
Riemann tensor $R_{\bf abc}^{\ \ \ \ \bf d}$ 
according to the convention 
\begin{equation} 
\nabla_{[\bf a}\nabla_{\bf b]}V^{\bf c} = \frac{1}{2} 
R_{\bf abd}^{\ \ \ \ \bf c}V^{\bf d}\ . \label{eq:rc} 
\end{equation} 
It follows that $R_{\bf abcd}:= R_{\bf abc}^{\ \ \ \ \bf e}
g_{\bf de}$ satisfies $R_{\bf abcd} = R_{\bf [ab][cd]}$, 
$R_{\bf abcd} = R_{\bf cdab}$, and $R_{\bf [abc]d}=0$. 
In the case of $RP^{n}$ with the natural metric 
\begin{equation} 
ds^{2}\ =\ \frac{8Z^{[a}dZ^{b]} 
Z_{[a}dZ_{b]}}{\lambda(Z^{c}Z_{c})^{2}}\ , 
\end{equation} 
where $Z^{a}$ are homogeneous coordinates and $\lambda$ is a 
scale factor (set to unity in the preceding analysis), 
the Riemann tensor is given by 
\begin{equation} 
R_{\bf abcd}\ =\ \lambda (g_{\bf ac}g_{\bf bd} - 
g_{\bf bc}g_{\bf ad})\ . \label{eq:RPcurv} 
\end{equation} 
\par 

Now we turn to consider projective transformations on $RP^{n}$. 
First we make a few general remarks about projective transformations 
on Riemannian manifolds \cite{tomonaga}. 
Suppose we have a Riemannian manifold with metric $g_{ab}$ and we 
consider the effect of dragging the metric along the integral 
curves of a vector field $\xi^{\bf a}$. For an infinitesimal 
transformation we have 
\begin{equation} 
g_{\bf ab}\ \rightarrow\ {\hat g}_{\bf ab}\ =\ g_{\bf ab} 
+ \epsilon {\cal L}_{\xi}g_{\bf ab}\ , \label{eq:inf} 
\end{equation} 
where ${\cal L}_{\xi}g_{\bf ab} = 2\nabla_{(\bf a}\xi_{\bf b)}$ is 
the Lie derivative of the metric ($\epsilon<<1$). The Levi-Civita 
connection ${\hat \nabla}_{\bf a}$ associated with 
${\hat g}_{\bf ab}$ is then defined by a tensor 
$Q^{\ \bf c}_{\bf ab}$, symmetric on its lower indices, such that 
\begin{equation} 
{\hat \nabla}_{\bf a}V^{\bf c}\ =\ \nabla_{\bf a}V^{\bf c} 
+ \epsilon Q^{\ \bf c}_{\bf ab}V^{\bf b}\ , \label{eq:conn} 
\end{equation} 
for an arbitrary vector field $V^{\bf a}$, and 
${\hat \nabla}_{\bf a}{\hat g}_{\bf bc}=0$. A familiar line of 
argument based on the fundamental theorem of Riemannian geometry 
then shows that 
\begin{equation} 
Q^{\ \bf c}_{\bf ab}\ =\ \frac{1}{2}{\hat g}^{\bf cd} \left( 
\nabla_{\bf a}{\hat g}_{\bf bd} + \nabla_{\bf b}{\hat g}_{\bf ad} 
+ \nabla_{\bf d}{\hat g}_{\bf ab} \right)\ , 
\end{equation} 
where ${\hat g}^{\bf ab}$ is the inverse of ${\hat g}_{\bf ab}$, 
given to first order in $\epsilon$ here by ${\hat g}^{\bf ab} = 
g^{\bf ab} - 2\epsilon \nabla^{(\bf a}\xi^{\bf b)}$. 
A short calculation making use of (\ref{eq:rc}) 
then shows that to first order we have: 
\begin{equation} 
Q^{\ \bf c}_{\bf ab}\ =\ \nabla_{(\bf a}\nabla_{\bf b)}\xi^{\bf c} 
+ R^{\bf c}_{\ \bf (ab)d}\xi^{\bf d}\ . \label{eq:Q} 
\end{equation} 
Now, suppose the vector field $U^{\bf a}$ satisfies the 
geodesic equation, which we write in the form $(U^{\bf a}
\nabla_{\bf a}U^{\bf [b})U^{\bf c]}=0$. It should be apparent that 
$U^{\bf a}$ is geodesic with respect to the transformed metric 
${\hat g}_{\bf ab}$ iff $(U^{\bf a}{\hat \nabla}_{\bf a}
U^{\bf [b})U^{\bf c]}=0$, or equivalently, $U^{\bf a}U^{\bf b}
Q^{\ \bf [c}_{\bf ab}U^{\bf d]}=0$. However, this relation is 
satisfied for {\it all} geodesic vector fields iff there exists a 
vector field $\phi_{\bf a}$ such that $Q^{\ \bf c}_{\bf ab} = 
\delta^{\bf c}_{(\bf a}\phi_{\bf b)}$. Hence we conclude that the 
vector field $\xi^{\bf a}$ is the generator of a projective 
transformation, mapping geodesics to geodesics, 
iff there exists a vector field $\phi_{\bf a}$ such that 
\begin{equation} 
\nabla_{(\bf a}\nabla_{\bf b)}\xi^{\bf c} + 
R^{\bf c}_{\ \bf (ab)d}\xi^{\bf d}\ =\ 
\delta^{\bf c}_{(\bf a}\phi_{\bf b)} \label{eq:projam} 
\end{equation} 
On the other hand, it follows as a direct consequence of 
(\ref{eq:rc}) and the identity $R_{\bf [abc]d}=0$ that for 
{\it any} vector field $\xi^{\bf a}$ we have 
\begin{equation} 
\nabla_{[\bf a}\nabla_{\bf b]}\xi^{\bf c} - 
R^{\bf c}_{\ \bf [ab]d}\xi^{\bf d}\ =\ 0\ . \label{eq:projam2}
\end{equation} 
Hence, combining (\ref{eq:projam}) and (\ref{eq:projam2}) we 
obtain 
\begin{equation} 
\nabla_{\bf a}\nabla_{\bf b}\xi^{\bf c} + 
R^{\bf c}_{\ \bf bad}\xi^{\bf d}\ =\ 
\delta^{\bf c}_{(\bf a}\phi_{\bf b)}\ , \label{eq:projam3} 
\end{equation} 
as a condition equivalent to (\ref{eq:projam}) for the existence 
of a projective transformation on the given manifold. This form of 
the condition is particularly useful for calculations 
(see, e.g., \cite{tomonaga}). \par 

In particular, if $\xi^{\bf a}$ is a Killing vector, so 
$\nabla_{\bf (a}\xi_{\bf b)}=0$, then (\ref{eq:projam3}) is 
automatically satisfied, with $\phi_{\bf a}=0$ since as a 
consequence of (\ref{eq:rc}) a Killing vector necessarily satisfies 
$\nabla_{\bf a}\nabla_{\bf b}\xi^{\bf c} + R^{\bf c}_{\ \bf bad}
\xi^{\bf c}=0$. Thus, a Killing symmetry generates a projective 
transformation. For example, in the case of $RP^{n}$ we have the 
symmetries of the projective orthogonal group. \par 

Now we derive an integrability condition for (\ref{eq:projam3}) 
that will be helpful in our analysis of the remaining projective 
transformations on $RP^{n}$, apart from those associated with 
Killing symmetries. Returning to the transformation 
(\ref{eq:inf}), we consider the curvature 
${\hat R}_{\bf abc}^{\ \ \ \ \bf d}$ associated with the metric 
${\hat g}_{\bf ab}$. Clearly, to first order in $\epsilon$ this 
is given by 
\begin{equation} 
{\hat R}_{\bf abc}^{\ \ \ \ \bf d}\ =\ 
R_{\bf abc}^{\ \ \ \ \bf d} + \epsilon {\cal L}_{\xi} 
R_{\bf abc}^{\ \ \ \ \bf d}\ . 
\end{equation} 
On the other hand, it is well known (see, e.g., \cite{lhpt}) 
that for a change 
of connection given by (\ref{eq:conn}) the corresponding change 
in the curvature is 
\begin{equation} 
\frac{1}{2}{\hat R}_{\bf abc}^{\ \ \ \ \bf d}\ =\ 
\frac{1}{2}R_{\bf abc}^{\ \ \ \ \bf d} + \epsilon \nabla_{[\bf a} 
Q^{\ \bf d}_{\bf b]c} + \epsilon^{2}Q^{\ \bf d}_{\bf r[a}
Q^{\ \bf r}_{\bf b]c}\ . 
\end{equation} 
Thus by consideration of terms to first order in $\epsilon$ we 
deduce that 
\begin{equation} 
\frac{1}{2}{\cal L}_{\xi}R^{\ \ \ \ \bf d}_{\bf abc}\ =\ 
\nabla_{\bf [a}Q^{\ \bf d}_{\bf b]c}\ , 
\end{equation} 
with $Q^{\ \bf c}_{\bf ab}$ given as in equation (\ref{eq:Q}). 
This relation holds for any infinitesimal transformation of the 
form (\ref{eq:inf}) generated by a given vector field $\xi^{\bf a}$. 
In the case of a projective transformation, for which $Q^{\ \bf c}
_{\bf ab}=\delta^{\bf c}_{(\bf a}\phi_{\bf b)}$, it follows, 
therefore, after some elementary rearrangement, that 
\begin{equation} 
{\cal L}_{\xi}R^{\ \ \ \ \bf d}_{\bf abc} + \delta^{\bf d}_{[\bf a}
\nabla_{\bf b]}\phi_{\bf c} - \nabla_{[\bf a}\phi_{\bf b]}
\delta^{\bf d}_{\bf c}\ =\ 0\ . \label{eq:integ} 
\end{equation} 
This is the desired integrability condition. In the case of a 
space of constant curvature we have (\ref{eq:RPcurv}) and hence 
\begin{equation} 
{\cal L}_{\xi}R^{\ \ \ \ \bf d}_{\bf abc}\ =\ 2\lambda
\nabla_{(\bf a}\xi_{\bf c)}\delta^{\bf d}_{\bf b} - 2\lambda 
\nabla_{(\bf b}\xi_{\bf c)}\delta^{\bf d}_{\bf a}\ . 
\end{equation} 
Inserting this relation into equation (\ref{eq:integ}) we obtain 
$A_{\bf ac}g_{\bf bd} - A_{\bf bc}g_{\bf ad} = 
\nabla_{[\bf a}\phi_{\bf b]}g_{\bf cd}$, where $A_{\bf ab} = 
2\lambda \nabla_{(\bf a}\xi_{\bf b)} - \nabla_{\bf a}\phi_{\bf b}$. 
This implies $A_{\bf ab}=0$ and 
$\nabla_{[\bf a}\xi_{\bf b]}=0$. For otherwise, we could 
contract $A_{\bf ab}$ with a vector $B^{\bf c}$ to obtain 
$C_{\bf a}g_{\bf bd}-C_{\bf b}g_{\bf ad} = \nabla_{[\bf a}
\phi_{\bf b]}B_{\bf d}$ where $C_{\bf a}=A_{\bf ac}B^{\bf c}$. 
But this would imply that $g_{\bf ab}$ is at most of rank two. 
It follows therefore that $\phi_{\bf a}$ is the gradient of a 
scalar, and that $\xi^{\bf a}$ is necessarily of the form 
\begin{equation} 
\xi^{\bf a}\ =\ \eta_{\bf a} + g^{\bf ab}\nabla_{\bf b}H\ , 
\label{eq:lemma} 
\end{equation} 
for some function $H(x)$ on the state manifold, 
where $\eta_{\bf a}$ is a Killing 
vector, and $\phi_{\bf a}=2\lambda\nabla_{\bf a}H$. Returning this 
information to equation (\ref{eq:projam}), we conclude that 
\begin{equation} 
\nabla_{(\bf a}\nabla_{\bf b}\nabla_{\bf c)}H\ =\ 
2\lambda g_{(\bf ab}\nabla_{\bf c)}H\ . \label{eq:hamf0} 
\end{equation} 
In other words, $K_{\bf ab}:= \nabla_{\bf a}\nabla_{\bf b}H 
- 2\lambda g_{\bf ab}H$ is a Killing tensor: $\nabla_{(\bf a}
K_{\bf bc)}=0$. However, this is the defining equation on 
$RP^{n}$ for observable functions, i.e., functions of the form 
\begin{equation} 
H(x)\ =\ \frac{H_{ab}Z^{a}Z^{b}}
{g_{cd}Z^{c}Z^{d}}\ , \label{eq:hamf}  
\end{equation} 
in terms of homogeneous coordinates on $RP^{n}$. Thus, we 
conclude that the generator of a projective transformation on the 
state space is given by the sum of a Killing vector and the 
gradient of an observable function. \par 

We remark, incidentally, that if the 
dimension $n$ of the manifold is finite, then we can form 
the trace of (\ref{eq:projam3}), which gives $\nabla_{\bf a}
(\nabla^{2}H) = \lambda(n+1)\nabla_{\bf a}H$, from which we 
conclude that up to an additive constant, $H(x)$ is an 
eigenfunction of the Laplacian operator, i.e., 
$\nabla^{2}H = \lambda(n+1)H$. The only global 
solutions of this equation on $RP^{n}$ are necessarily of the 
form (\ref{eq:hamf}). Such functions also satisfy 
(\ref{eq:hamf0}). In 
the case of infinite dimension the trace is not necessarily a 
valid operation, and we lose the representation of observable 
functions as eigenfunctions of the Laplacian. On the other hand, 
our characterisation of projective transformations as vector 
fields of the form (\ref{eq:lemma}), given by the sum of a Killing 
vector and the gradient of a scalar $H(x)$ satisfying 
(\ref{eq:hamf0}) and (\ref{eq:hamf}), is entirely general, valid in 
both finite and infinite dimensions. Setting $\lambda=1$, we see 
therefore that thermal states are characterised by vector fields 
on $RP^{n}$ of the form $\xi^{\bf a}=g^{\bf ab}\nabla_{\bf b}H$, 
where $H(x)$ satisfies $\nabla_{(\bf a}\nabla_{\bf b}
\nabla_{\bf c)}H = 2 g_{(\bf ab}\nabla_{\bf c)}H$. \par 

It is worthwhile noting that among the trajectories generated by 
the Hamiltonian gradient flow, the thermal trajectory is singled 
out on account of the special initial state corresponding to 
$\beta=0$. This state is, of course, completely determined 
by the specification of the Hamiltonian function. In particular, 
for a given Hamiltonian $H_{ab}$ there are in general $n+1$ fixed 
points in the state manifold. The initial state for the thermal 
trajectory is then uniquely determined by the condition that it 
is equidistant from these $n+1$ fixed points. \par 

\section{Quantum Statistical Mechanics} 

\subsection{The quantum phase space} 

In the algebraic approach to quantum theory one 
generally works with the totality of mixed states, while regarding 
pure states as extremal elements of this convex set \cite{hhw}, and 
in the Heisenberg picture for dynamics. The approach we adopt here, 
however, will focus primarily on the system of pure 
states, and for dynamics we shall work in the 
Schr\"odinger picture. This is consistent with the point of view we 
put forward earlier in Section 2 in connection with the discussion 
of classical thermal states. \par 

Given a complex Hilbert space ${\cal H}_{\bf C}^{n+1}$ with complex 
coordinates $Z^{\alpha}$ representing pure quantum states, 
we identify the state vector $Z^{\alpha}$ with its complex 
multiples $\lambda Z^{\alpha}$, $\lambda\in{\bf C}-\{0\}$, to obtain 
the complex projective space $CP^{n}$. Here we use Greek indices 
($\alpha=0,1,\cdots,n$) for vectors in ${\cal H}_{\bf C}^{n+1}$. 
Following the line of argument indicated earlier in the case of 
the thermal state space, we regard $CP^{n}$ as the true `state 
space' of quantum mechanics. The status of density matrices will be 
discussed shortly. Our goal now is to build up the necessary 
projective geometric machinery appropriate for the consideration of 
quantum mechanics. \par 

Suppose we regard the state vector $Z^{\alpha}$ as representing 
homogeneous coordinates on the projective Hilbert space. The complex 
conjugation of the state vector $Z^{\alpha}$ is written 
${\bar Z}_{\alpha}$, with the corresponding Hermitian inner product 
$Z^{\alpha}{\bar Z}_{\alpha}$. The complex conjugate of a point 
$P^{\alpha}$ in $CP^{n}$ is the hyperplane (prime) 
${\bar P}_{\alpha}Z^{\alpha}=0$. The points on this plane are 
the states that are orthogonal to the original state $P^{\alpha}$, and 
we denote this $\neg P^{\alpha}$ in the sense of the probability 
rules discussed above. Thus, 
$CP^{n}$ is equipped with a Hermitian correlation, i.e., a 
complex conjugation operation that maps points to hyperplanes of 
codimension one, that is, $CP^{n-1}$. \par 

Distinct points $X^{\alpha}$ 
and $Y^{\alpha}$ are joined by a complex projective line 
$L^{\alpha\beta} = X^{[\alpha}Y^{\beta]}$, representing the various 
complex, quantum mechanical superpositions of the original two 
states. 
The quantum mechanical transition probability between two states 
$X^{\alpha}$ and $Y^{\alpha}$ is then given by the cross ratio 
\begin{equation} 
\kappa\ =\ \frac{X^{\alpha}{\bar Y}_{\alpha}Y^{\beta}
{\bar X}_{\beta}}{X^{\gamma}{\bar X}_{\gamma}Y^{\delta}
{\bar Y}_{\delta}}\ . 
\end{equation} 
More precisely, we recall that if the system is in the state 
$Y^{\alpha}$ and a measurement is made to see if the system is in 
the state $X^{\alpha}$ (corresponding to a measurement of the 
observable represented by the projection operator $X^{\alpha} 
{\bar X}_{\beta}/X^{\gamma}{\bar X}_{\gamma}$), then $\kappa$ is 
the probability that the result of the measurement is affirmative. 
If we set $\kappa=\cos^{2}(\theta/2)$ and set $X^{\alpha} = 
Z^{\alpha}$ and $Y^{\alpha} = Z^{\alpha}+dZ^{\alpha}$, retaining 
terms to second order, we recover the natural unitary-invariant 
metric on complex projective space, known as the Fubini-Study 
metric \cite{kn}, given by 
\begin{equation} 
ds^{2}\ =\ 8(Z^{\alpha}{\bar Z}_{\alpha})^{-2} 
Z^{[\alpha}dZ^{\beta]}{\bar Z}_{[\alpha}d{\bar Z}_{\beta]}\ . 
\label{eq:FS} 
\end{equation} 

Now suppose we write $H^{\alpha}_{\beta}$ for the Hamiltonian 
operator, assumed Hermitian. Then for the Schr\"odinger equation 
we have 
\begin{equation} 
\frac{dZ^{\alpha}}{dt}\ =\ iH^{\alpha}_{\beta}Z^{\beta}\ . 
\end{equation} 
However, we are more interested in the {\it projective} 
Schr\"odinger equation, given by 
\begin{equation} 
Z^{[\alpha}dZ^{\beta]}\ =\ iZ^{[\alpha}H^{\beta]}_{\gamma}
Z^{\gamma}dt\ , \label{eq:scheq} 
\end{equation} 
which eliminates the superfluous degree of freedom associated with 
the direction of $Z^{\alpha}$. This equation is well defined on the 
projective state space. Insertion of (\ref{eq:scheq}) into 
(\ref{eq:FS}) gives  
\begin{equation} 
ds^{2}\ =\ 4\langle{\tilde H}^{2}\rangle dt^{2}\ , 
\end{equation} 
which verifies that the velocity of a state along its trajectory 
in state space is $ds/dt=2\Delta H$, where $\Delta H = 
\langle{\tilde H}^{2}\rangle^{1/2}$ is the energy uncertainty 
\cite{aa}. \par 

An alternative way of looking at this structure is to regard the 
state manifold $CP^{n}$ as a real manifold 
${\cal M}^{2n}$ of dimension $2n$, equipped with a Riemannian 
metric $g_{\bf ab}$ and a compatible complex structure $\Omega_{\bf 
ab}$. Here again we use bold indices for tensorial operations in 
the tangent space of ${\cal M}^{2n}$. In this 
formulation the Schr\"odinger equation takes the form 
of a flow $\xi^{\bf a}$ on ${\cal M}^{2n}$ given by 
\begin{equation} 
\xi^{\bf a}\ =\ \Omega^{\bf ab}\nabla_{\bf b}H\ . 
\label{eq:sch} 
\end{equation} 
Here $H$ is the real function on ${\cal M}^{2n}$ given at each point 
by the expectation of the Hamiltonian operator at that state. 
It is interesting to note 
that in the quantum mechanical case the dynamical trajectory is 
given by a Hamiltonian {\it symplectic} flow in contrast to the 
case of statistical mechanics where the relevant flow is given by 
a Hamiltonian {\it gradient} flow for the thermal trajectories. 
Furthermore, we note that in the case of quantum mechanics the 
flow is necessarily a 
Killing field, satisfying $\nabla_{(\bf a}\xi_{\bf b)}=0$, 
where $\xi_{\bf a}=g_{\bf ab}\xi^{\bf b}$. In other words, the 
isometries of the Fubini-Study metric on $CP^{n}$ can be lifted 
to ${\cal H}^{n+1}_{\bf C}$ to yield unitary transformations. In the 
statistical mechanical case, on the other hand, the general 
projective transformation of the manifold includes both Killing 
vectors and Hamiltonian gradient flows, but we exclude the former. 
In both quantum mechanics and 
statistical mechanics the fixed points of the Hamiltonian operator 
play a pivotal role. \par 

\subsection{The quantum thermal states} 

In Section 4.A above we have indicated two useful ways of 
looking at the quantum state space. The approach via complex
homogeneous coordinates is appropriate when the {\it global} 
algebraic geometry of the state manifold is of interest. 
The differential geometric approach is useful when {\it local} 
properties of the state manifold are considered \cite{lph}. There 
is a third, equally important approach, however, which we shall 
call the `real' approach \cite{dblh0,dblh2,a-s} to quantum 
mechanics. In this case we regard the Hilbert space of quantum 
mechanics as a real 
vector space of dimension $2n+2$. The importance of the real 
approach is that it links up directly with modern notions from 
probability theory and statistics. It should not therefore be 
surprising that this approach is also useful in considering the quantum 
dynamics of thermal states, and indeed in considering the relation 
between classical and quantum statistical mechanics. \par 

Let us write $\xi^{a}$ $(a=1, 2, \cdots, 2n+2)$ for the real 
Hilbert space coordinates corresponding to a quantum mechanical 
state vector. The other ingredients we have at our disposal 
are the metric $g_{ab}$ on ${\cal H}^{2n+2}$ and a compatible 
complex structure tensor $J^{b}_{a}$. Therefore, we proceed 
in two stages. First we regard the state vector $\xi^{a}$ as 
homogeneous coordinates for a real projective space $RP^{2n+1}$ of 
one dimension lower. This, of course, corresponds to factoring out 
by the freedom $\xi^{a}\rightarrow\lambda\xi^{a}$ with $\lambda$ 
real and nonvanishing. Then there is another more subtle freedom, 
corresponding to factoring out by the `phase' freedom. This is 
given by the transformation 
\begin{equation} 
\xi^{a}\ \rightarrow\ \frac{1}{2}e^{i\phi}
(\xi^{a}+iJ^{a}_{b}\xi^{b}) + \frac{1}{2}e^{-i\phi} 
(\xi^{a}-iJ^{a}_{b}\xi^{b})\ , 
\end{equation} 
where $J^{a}_{b}$ is the complex structure, and $\phi$ is a 
phase factor. In this way we obtain a map $RP^{2n+1}\rightarrow 
CP^{n}$, which is related to the Hopf map 
$S^{2n+1}\rightarrow CP^{n}$, and indeed we have the diagram 
\begin{eqnarray} 
\begin{array}{ccccc} 
{\cal H}^{2n+2} & \longrightarrow & S^{2n+1} & &  \\ 
 & \searrow & \downarrow & \searrow & \\ 
 & & RP^{2n+1} & \longrightarrow & CP^{n} 
\end{array} 
\end{eqnarray} 
showing the relation between these various spaces. The point of 
this line of reasoning is that if we take the classical statistical 
argument based on the state space $RP^{2n+1}$, and introduce a 
complex structure on the real Hilbert space ${\cal H}^{2n+2}$, 
then the analogous relations for quantum statistical mechanics, 
and moreover, quantum dynamics of the equilibrium thermal states 
can be developed by studying the relation between $RP^{2n+1}$ and 
$CP^{n}$. \par 

Now we consider more explicitly the characterisation of thermal 
states in the context of quantum `thermodynamics'. There are a 
number of features of the Hilbert space based description of 
conventional statistical mechanics that carry through to the quantum 
regime, though there are some new features as well. In particular, 
if we start with the real Hilbert space ${\cal H}^{2n+2}$ and the 
associated state space $RP^{2n+1}$ obtained when we neglect the 
complex structure on ${\cal H}^{2n+2}$ given by quantum theory, 
then we can apply the theory developed earlier to define thermal 
trajectories in the space $RP^{2n+1}$. The main 
difference arising between the classical and quantum cases is that 
in the quantum theory the Hamiltonian operator $H_{ab}$ satisfies 
the Hermitian property 
\begin{equation} 
H_{ab}J^{a}_{c}J^{b}_{d}\ =\ H_{cd}\ , 
\end{equation} 
which ensures that it has at most $n+1$ distinct 
eigenvalues, unlike the classical situation where there could be 
$2n+2$. The thermal equation giving the development of $\psi^{a}
(\beta)$ as the inverse temperature $\beta$ is changed, is again 
given by 
\begin{equation} 
\frac{d\psi^{a}(\beta)}{d\beta}\ =\ \frac{1}{2} 
{\tilde H}^{a}_{b}\psi^{b}(\beta)\ , \label{eq:qther} 
\end{equation} 
where ${\tilde H}_{ab} = H_{ab} - g_{ab}H_{cd}\psi^{c}\psi^{d}
/g_{ef}\psi^{e}\psi^{f}$. \par 

Another feature that distinguishes the quantum theory is that in 
general there is a multiplicity of infinite temperature $(\beta=0)$ 
states, corresponding to different values of the `phases' 
associated with the state. Topologically, this implies that there 
is a torus $T^{n}=(S^{1})^{n}$ in $CP^{n}$, the points of which 
correspond to the $\beta=0$ state. Thus, for a 
specified infinite temperature state, i.e., for a given point on 
$T^{n}$, we shall call the solution of (\ref{eq:qther}) a 
`primitive' thermal state parameterised by a set of phases. In 
other words, there is a multiplicity of primitive thermal 
trajectories, each corresponding to a choice of infinite 
temperature state. In the case of a two state system (e.g., a 
spin one-half particle) the two energy eigenstates correspond to 
the north and south poles of the state space $CP^{1}\sim S^{2}$, 
and the infinite temperature states are given 
by points on the equator taking different values of the phases. The 
thermal trajectories are then given by the geodesic segments that 
join the equator to one of the poles. This example is studied in 
more detail in Section 4.D. \par 

For many applications we have to consider ensembles of particles 
for which the starting point of the thermal trajectory is effectively 
random. In other words, although each particle in the ensemble has 
a definite thermal trajectory, 
when we form expectations based on the behaviour of 
the ensemble we have to average over the manifold of infinite 
temperature states, i.e., average over the random phase factors. 
This leads to the well known density matrix based description of 
thermal states, as we discuss below. Thus once again the density 
matrix formulation emerges as an essentially secondary 
construction, although of course it remains essential for many 
practical applications. \par 

Another characteristic of quantum thermal dynamics is that the 
thermal trajectories have a time evolution, given by the 
Schr\"odinger equation. Written in terms of real state vectors this 
evolution is given by 
\begin{equation} 
\psi^{a}_{\phi}(t,\beta)\ =\ \exp(-tJ^{a}_{b}{\tilde H}^{b}_{c})
\psi^{c}_{\phi}(0,\beta)\ , 
\end{equation} 
for each thermal trajectory, where the variables collectively 
denoted $\phi$ label the $\beta=0$ states. Thus the primitive 
thermal states are pure states, subject to the usual evolutionary 
laws. \par 

In our earlier discussion of classical thermal systems, we 
introduced the idea of a random state. This notion is also 
useful in a quantum mechanical context. For example, a random state 
${\bf \Psi}^{a}$ can be used to represent the outcome of a 
measurement process, or to describe the statistics of an ensemble. 
In particular, as we 
indicated earlier, in the case of a quantum thermal state we 
want to take an ensemble average over all possible infinite 
temperature states to describe the observed features of 
ensembles. This point will now be explored further. \par 

Suppose we denote by ${\bf \Phi}$ the system of random phases 
labelling the $\beta=0$ states for a given quantum ensemble. 
The thermal state $\psi^{a}_{\bf \Phi}$ is thus itself to be viewed 
as a random state, since ${\bf \Phi}$ is random. Then if $X_{ab}$ is a 
typical observable, satisfying the Hermitian property, we can write 
$E_{\bf \Phi}[X] = X_{ab}\psi^{a}_{\bf \Phi}\psi^{b}_{\bf \Phi}$ for 
the conditional expectation of $X_{ab}$, given ${\bf \Phi}$. This is 
in keeping with the line of argument introduced earlier, for 
classical thermal systems. The ensemble average $E[X]$ is then given 
by the unconditional expectation 
\begin{equation} 
E[E_{\bf \Phi}[X]]\ =\ X_{ab}\rho^{ab}\ , 
\end{equation} 
where 
\begin{equation} 
\rho^{ab}(\beta)\ =\ E[\psi^{a}_{\bf \Phi}\psi^{b}_{\bf \Phi}] 
\label{eq:rho} 
\end{equation} 
is the Hermitian density matrix characterising the state of the 
ensemble. Here the unconditional expectation $E[-]$ averages 
over the phase variables appearing in random state 
$\psi^{a}_{\bf \Phi}(0,\beta)$. In this connection it is worth 
noting that the infinite temperature states have the property 
that the ensemble average in formula (\ref{eq:rho}) is 
proportional to the metric for $\beta=0$, that is, 
$\rho^{ab}(0)=\frac{1}{2n+2}g^{ab}$, where $2n+2$ is the 
dimension of the real Hilbert space. \par 

It is interesting to take note of the topological 
characteristics of the infinite temperature state manifold. As we 
already remarked, this manifold can be viewed as an $n$-torus 
$T^{n}$ sitting in the state manifold $CP^{n}$. This can be 
seen as follows. First, recall the fact that the $\beta=0$ states 
are equidistant from the fixed points $u^{\alpha}_{k}$ of the 
Hamiltonian flow in $CP^{n}$, and that these fixed points form a 
regular simplex. Thus, we would like to find the locus of 
points $Z^{\alpha}$ such that the distances 
between $Z^{\alpha}$ and $u^{\alpha}_{k}$ are equal for all $k$. 
Since the points $u^{\alpha}_{k}$ are vertices of a regular 
simplex, we can choose coordinates, without loss of generality, 
such that 
$u^{\alpha}_{0}=(1,0,0,\cdots)$, $u^{\alpha}_{1}=(0,1,0,\cdots)$, 
and so on. Thus, by writing the equidistance condition in terms 
of the cross ratio explicitly, we find that the locus of 
$Z^{\alpha} = (Z^{0}, Z^{1}, \cdots, Z^{n})$ is given by the 
simultaneous solution of the following system of equations: 
$Z^{0}{\bar Z}_{0} = Z^{1}{\bar Z}_{1}, 
Z^{1}{\bar Z}_{1} = Z^{2}{\bar Z}_{2}, \cdots, 
Z^{n}{\bar Z}_{n} = Z^{0}{\bar Z}_{0}$. 
We thus deduce that the solution takes the form 
\begin{equation} 
Z^{\alpha}\ =\ (e^{i\phi_{0}}, e^{i\phi_{1}}, \cdots, 
e^{i\phi_{n}})\ , 
\end{equation} 
for some phase variables $\phi_{j}$. This implies that the 
topology of the infinite temperature states manifold indeed has 
a structure of an $n$-torus $T^{n}=(S^{1})^{n}$ (one of the 
$n+1$ phase variables $\phi_{j}$ is redundant, and can be 
scaled away), as claimed earlier. In summary, the fundamental 
torus $T^{n}\subset CP^{n}$ associated with a given 
Hamiltonian function consists of all those states 
that are equally distant to the eigenstates of the specified 
Hamiltonian with respect to the Fubini-Study metric on 
$CP^{n}$. The significance of $T^{n}$ is that it embodies a 
geometrical representation of the concept of the extraneous 
phases over which one has to integrate for many purposes 
in thermal physics. \par 

\subsection{The equilibrium condition}  

The property that the density matrix 
$\rho^{ab}(0)$ at infinite temperature is proportional to the 
metric is, in fact, closely related to the KMS-condition 
\cite{hhw,b-r}, which provides an alternative characterisation 
of thermal equilibrium states. It is therefore interesting to 
see how the KMS construction fits into the present description 
of thermal phenomena, and here we shall briefly develop some of 
the relevant 
ideas. First, recall that a general state, or density 
matrix, can be regarded as a semidefinite map $\rho(A)$ from 
Hermitian operators $A$ to the real numbers given by $\rho(A) = 
{\rm Tr}(\rho A)$, satisfying $\rho(X{\bar X})\geq0$ for all 
$X$, and $\rho(g)=1$, where $g$ is the identity operator. Now 
the trace operation has the property ${\rm Tr}(AB) = 
{\rm Tr}(BA)$. For a general state $\rho$, on the other hand, 
clearly we do not have $\rho(AB)=\rho(BA)$. However, the thermal 
equilibrium states $\rho_{\beta}(-)$ are characterised by a 
slightly weakened form of commutation relation, which is the 
KMS condition. \par 

Suppose for a Hermitian operator $A$ we write $\tau_{t}A$ for the 
unitary action of the one-parameter group of time 
translations given 
by $\tau_{t}A = e^{itH}Ae^{-itH}$. Then, we can extend this 
definition to complex time by writing $\tau_{t+i\beta}A = 
e^{i(t+i\beta)H}Ae^{-i(t+i\beta)H}$. The KMS condition on a 
state $\rho$ is that for all Hermitian operators $A$ and $B$ 
it should satisfy 
\begin{equation} 
\rho(\tau_{t}A B)\ =\ \rho(B \tau_{t+i\beta}A)\ . 
\end{equation} 
It is not difficult to see that in finite dimensions this 
implies that $\rho$ is the thermal state $\rho_{\beta}$ 
discussed above, and that commutivity holds for infinite 
temperature. \par 

Now let us look at these relations from a real point of view. 
First we need to consider the action of the one-parameter 
group of time translations. Suppose we have a real Hilbert 
space of dimension $2n+2$, with typical elements 
$\xi^{a}, \eta^{a} \in {\cal H}$. On ${\cal H}$ we also have the 
metric $g_{ab}$ and the complex structure $J^{b}_{a}$, satisfying 
$J^{c}_{a}J^{b}_{c}=-\delta^{b}_{a}$ and $J^{c}_{a}J^{d}_{b}g_{cd} 
= g_{ab}$. The tensor $\Omega_{ab}=g_{ac}J^{c}_{b}$ is thus 
antisymmetric, and defines a symplectic structure on ${\cal H}$. 
With these ingredients at hand, we can define the action of the 
orthogonal and unitary groups on ${\cal H}$. \par 

For a typical element $\xi^{a}\in{\cal H}$ the orthogonal group 
$O(2n+2)$ consists of transformations $\xi^{a}\rightarrow 
O^{a}_{b}\xi^{b}$, satisfying $O^{c}_{a}O^{d}_{b}g_{cd}=g_{ab}$. 
The unitary group $U(n+1)$ can then be regarded as a subgroup of 
$O(2n+2)$, given by transformations $\xi^{a}\rightarrow 
U^{a}_{b}\xi^{b}$ satisfying 
\begin{equation} 
U^{c}_{a}U^{d}_{b}g_{cd}\ =\ g_{ab}\ , \ \ \ 
U^{c}_{a}U^{d}_{b}\Omega_{cd}\ =\ \Omega_{ab}\ . \label{eq:unit2} 
\end{equation} 
In other words, transformations that preserve both the metric and 
the symplectic structure on ${\cal H}$. For typical real elements 
$\xi^{a}, \eta^{a} \in {\cal H}$ their ordinary real Hilbert space 
inner product is, of course, given by $\eta^{a}g_{ab}\xi^{b}$, 
which is invariant if $\xi^{a}$ and $\eta^{a}$ are subject to 
orthogonal transformations. On the other hand, the standard Dirac 
product $\langle\eta|\xi\rangle$ between two real vectors $\xi^{a}$ 
and $\eta^{a}$, given by 
\begin{equation} 
\langle\eta|\xi\rangle\ =\ \frac{1}{2}\eta^{a}
(g_{ab}-i\Omega_{ab})\xi^{b}\ , \label{eq:dirac} 
\end{equation} 
is invariant only under unitary transformations. This should be 
clear from the need to preserve the real and imaginary part of 
(\ref{eq:dirac}) separately, which implies (\ref{eq:unit2}). \par 

If $O^{b}_{a}$ is an orthogonal transformation, then the 
corresponding inverse element ${\bar O}^{b}_{a}$ is given by the 
transpose ${\bar O}^{b}_{a}=g^{bc}g_{ad}O^{d}_{c}$, which 
satisfies $O^{a}_{b}{\bar O}^{b}_{c}=\delta^{a}_{c}$. In the case 
of a unitary transformation we can, without ambiguity, use the 
notation ${\bar U}^{b}_{a} = g^{bc}g_{ad}U^{d}_{c}$ for the 
conjugate transformation, from which it follows that 
\begin{equation} 
U^{a}_{b}{\bar U}^{b}_{c}\ =\ \delta^{a}_{c}\ , \ \ \ 
U^{a}_{b}J^{b}_{c}{\bar U}^{c}_{d}\ =\ \delta^{a}_{d} 
\label{eq:unit4} 
\end{equation} 
are equivalent to (\ref{eq:unit2}) in 
characterising unitary transformations. In particular, 
(\ref{eq:unit4}) can be viewed as saying that 
the unitary transformations on ${\cal H}$ are orthogonal 
transformations that also preserve the complex structure. \par 

In equation (\ref{eq:unit4}) we see the action of the unitary 
group on the complex structure tensor. More generally, for a typical 
multi-index tensor $A^{a}_{bc}$ we define the action of the 
unitary group by $A^{a}_{bc}\rightarrow U^{a}_{a'}A^{a'}_{b'c'} 
{\bar U}^{b'}_{b}{\bar U}^{c'}_{c}$. Here the primed indices merely 
serve to increase the size of the standard alphabet. \par 

Now suppose we consider one-parameter subgroups of orthogonal 
group, continuous with the identity. 
Such transformations are of the form $O^{b}_{a} = 
\exp[tM_{ac}g^{bc}]$ where the tensor $M_{ab}$ is antisymmetric. 
For a one-parameter family of unitary transformations we have to 
specialise further, and require $M_{ab}$ to be of the form 
$M_{ab}=\Omega_{ac}H^{c}_{b}$, where $H_{ab}$ is symmetric and 
Hermitian. Indeed, if $H_{ab}$ is symmetric, then a necessary 
and sufficient condition that it should be Hermitian is that 
$\Omega_{ac}H^{c}_{b}$ is antisymmetric. It follows that the 
general one-parameter group of unitary transformations, continuous 
with the identity, can be written 
\begin{equation} 
\xi^{a}\ \rightarrow\ \xi^{a}_{t} = 
\exp\left[ tJ^{a}_{b}H^{b}_{c} 
\right] \xi^{c}\ . 
\end{equation} 
If $H_{ab}$ is the Hamiltonian, then $\xi^{a}_{t}$ is the 
Schr\"odinger evolution generated from the given initial state. 
The point of view here continues to be purely `real' in the 
sense that the complex structure tensor $J^{a}_{b}$ is playing 
the role of the factor of `$i$' in the conventional expression 
of unitary evolution given in the Dirac notation by $|\xi\rangle 
\rightarrow e^{itH}|\xi\rangle$. Our goal is to formulate a 
similar geometrisation for the KMS condition. \par 

For this purpose we need to consider the changes implied 
for the picture noted above when we go to the Heisenberg 
representation. Suppose $A_{ab}$ is an observable (symmetric, 
Hermitian operator) and we consider the evolution of its expectation 
$A_{ab}\xi^{a}\xi^{b}$ under the action of the unitary 
transformation $\xi^{a}\rightarrow U^{a}_{b}\xi^{b}$ with 
$U^{a}_{b}=\exp[tJ^{a}_{c}H^{c}_{b}]$. The Heisenberg picture is 
obtained if we let $\xi^{a}$ be stationary, and let $A_{ab}$ evolve 
according to the scheme 
\begin{equation} 
A_{ab}\ \rightarrow\ U^{c}_{a}U^{d}_{b}A_{cd}\ . 
\end{equation} 
It will be appreciated that the evolution of $A_{ab}$ in the 
Heisenberg representation is contragradient to the `natural' action 
of the unitary group $A_{ab}\rightarrow 
{\bar U}^{a'}_{a}{\bar U}^{b'}_{b} A_{a'b'}$ 
discussed earlier, since the natural action of the 
unitary group is defined to be the action on $A_{ab}$ that preserves 
$A_{ab}\xi^{a}\xi^{b}$ when $\xi^{a}$ is evolved in the 
Schr\"odinger representation. By the same token, in the Heisenberg 
representation the action of the time evolution operator on an 
observable, represented in `operator' form $A^{b}_{a}$ (rather than 
$A_{ab}$) is given by 
\begin{equation} 
A^{b}_{a}\ \rightarrow\ U^{a'}_{a}A^{b'}_{a'}{\bar U}^{b}_{b'}\ , 
\end{equation} 
where $U^{a'}_{a}=\exp[tJ^{a'}_{b}H^{b}_{a}]$ and 
${\bar U}^{a'}_{a}=\exp[-tJ^{a'}_{b}H^{b}_{a}]$. \par 

Now, finally, we are in a position to address the KMS condition. 
First we note that for a pair of observables 
$A_{ab}$ and $B_{ab}$, their quantum mechanical `Dirac' product 
$C=AB$ is given by 
\begin{equation} 
C_{ab}\ =\ A_{c(a}\Delta^{cd}B_{b)d}\ , 
\end{equation} 
where $\Delta^{ab} = \frac{1}{2}(g^{ab}-i\Omega^{ab})$. Clearly, 
$C_{ab}$ is a complex tensor: its real and imaginary parts are given 
respectively by the Jordan product $\frac{1}{2}(AB+BA)$ and the 
commutator $\frac{1}{2}i(AB-BA)$. In particular, we have 
\begin{equation} 
C_{ab}\ =\ \frac{1}{2}A^{c}_{(a}B^{d}_{b)}g_{cd} - \frac{1}{2}i 
A^{c}_{(a}B^{d}_{b)}\Omega_{cd}\ . 
\end{equation} 
\par 

It follows that if $\rho_{ab}$ is a density matrix, the expectation 
$\rho(AB)$ in the corresponding state is given by 
$\rho(AB) = \rho^{a}_{b}A^{b}_{c}\Delta^{c}_{d}B^{d}_{a}$, or 
equivalently 
\begin{equation} 
\rho(AB)\ =\ \frac{1}{2}\rho_{ab}A^{a}_{c}B^{bc} - \frac{1}{2}i 
\rho_{ab}A^{a}_{c}B^{b}_{d}\Omega^{cd}\ . 
\end{equation} 
As we noted earlier, the action of the one-parameter group 
$\tau_{t}$ on $A$ is given by 
\begin{equation} 
A^{b}_{a}\ \rightarrow\ \tau_{t}A^{b}_{a}\ :=\ 
\exp\left[ tJ^{c}_{a}H^{a'}_{c}\right] A^{b'}_{a'} 
\exp\left[ -tJ^{d}_{b'}H^{b}_{d}\right]\ . 
\end{equation} 
The KMS complexification of this action, corresponding to replacing 
$t$ with $t+i\beta$, replaces $tJ^{b}_{a}$ with $tJ^{b}_{a} - 
\beta \delta^{b}_{a}$, and we have 
\begin{equation} 
\tau_{t+i\beta}A^{b}_{a}\ :=\ \exp\left[ tJ^{c}_{a}H^{a'}_{c} - 
\beta H^{a'}_{a}\right] A^{b'}_{a'} \exp\left[ -tJ^{d}_{b'}
H^{b}_{d} + \beta H^{b}_{b'}\right]\ . \label{eq:im-t} 
\end{equation} 
It is then straightforward to verify that in the case of 
a thermal state, for which 
\begin{equation} 
\rho^{b}_{a}\ =\ \frac{\exp[-\beta H^{b}_{a}]}
{\delta^{c}_{d}\exp[-\beta H^{d}_{c}]}\ , 
\end{equation} 
the KMS condition $\rho_{\beta}(\tau_{t}A B) = 
\rho_{\beta}(B \tau_{t+i\beta}A)$ is indeed satisfied. \par 

It might be argued that the complexification $t\rightarrow t+i\beta$ 
has an artificial character when seen from a real point of 
view, since it involves an operator transformation of the form 
$t\delta^{b}_{a} \rightarrow t\delta^{b}_{a} + \beta J^{b}_{a}$. 
Moreover, equation (\ref{eq:im-t}) takes one outside the category 
of symmetric tensors. Since we would like to argue that the `real' 
approach to quantum theory acts as the natural bridge between quantum 
dynamics, on the one hand, and modern statistical theory and hence 
thermal physics, on the other hand, we are thus led to the 
negative conclusion that the KMS condition, despite its historical 
significance in the development of modern thermal physics, should not 
be regarded as fundamental. This is in keeping with our emphasis on 
so-called primitive thermal states, which play a 
significant role even before the consideration of ensemble
behaviour. \par 

\subsection{Phase space dynamics and temperature} 

In the foregoing formulation, we have adopted the view point that, 
given a system in heat bath with inverse temperature $\beta$, we 
let the equilibrium thermal states evolve quantum mechanically under 
the influence of the Hamiltonian. We have also 
considered the case of an ensemble of particles in a thermalised box, 
whereby while each particle takes a definite energy value, the 
probability law of the energy for a random element of the ensemble 
being given by the Boltzmann distribution. In either case, the value 
of $\beta$ is taken as an `input' variable that specifies the thermal 
state. That is to say, we have adopted the standard canonical 
ensemble of distributions for which temperature is operationally 
defined. \par 

This point of view can, to some extent, be inverted by shifting 
gears from the standard canonical 
description to the microcanonical ensemble. In this case, the 
essential equivalence of the two approaches is typically 
recognised only for large systems (see, e.g., \cite{ruell}). The 
dynamical formulation we provide here will be useful, in 
particular, for small systems such as those on a quantum or 
mesoscopic scale. \par 

First, we regard as given a Hamiltonian function $H$ defined on the 
quantum mechanical state space $CP^{n}$, which is viewed as a real 
manifold of dimension $2n$. We then foliate the real state 
manifold with constant energy surfaces, given by $H(x) = E$. The 
volume ${\cal V}(E)$ of such a surface ${\cal E}_{E}^{2n-1}$, given 
by 
\begin{equation} 
{\cal V}(E)\ =\ \int_{{\cal E}_{E}^{2n-1}} 
d\sigma^{a} \frac{\nabla_{a}H}{\|{\rm grad}H\|}\ , 
\end{equation} 
then tells us the number of microscopic states having the energy 
$E$. Here, we have written 
$d\sigma^{a} = g^{ab}\epsilon_{bc\cdots d}dx^{c}\cdots dx^{d}$ 
for the natural vector-valued ($2n-1$)-form on $CP^{n}$ (viewed as 
a real manifold). Therefore, using the Boltzmann relation, the 
entropy $S(E)$ associated with the energy surface ${\cal E}_{E}$ 
is given by 
$S(E) = \ln {\cal V}(E)$. As a result, the inverse temperature can 
be calculated from the usual prescription 
\begin{equation} 
\beta\ =\ \frac{dS(E)}{dE}\ . \label{eq:temp} 
\end{equation} 
In this way, we can calculate the temperature of the system 
directly from the given quantum mechanical dynamics of the system, 
or equivalently, given the inter-relationship of the 
energy surfaces.  \par 

The construction just noted for obtaining temperature from the 
underlying phase space dynamics is well known in classical 
statistical mechanics (see, e.g., \cite{thompson}). The novelty here, 
however, is to regard the quantum mechanical state space $CP^{n}$ as 
a real $2n$-dimensional manifold playing the role of the quantum 
mechanical phase space, hence allowing us to obtain the temperature 
for quantum mechanical dynamical systems in the case of the 
microcanonical ensemble. Based upon this formulation, we can also 
investigate the possibility of various generalisations of the 
temperature, when the underlying quantum mechanical dynamics is 
modified. There are a number of distinct generalisations that can 
be pursued in this context. \par 

First, we can replace the Hamiltonian function defined on the state 
manifold by a more general observable. In this case, the underlying 
theory would correspond to the Kibble-type theories of 
nonlinear quantum mechanics, for which the above prescription would 
provide the temperature in the case of a nonlinear dynamical system 
of this sort. Secondly, we can consider replacing 
the Fubini-Study metric on the state space 
by a more general metric. Alternatively, a more radical extension 
is obtained by replacing the state space manifold itself with a general 
K\"ahler manifold. In such generalisations the notion of particle 
states may no longer survive. On the face of it, this might appear 
to be an undesirable feature. However, our view is that at high 
energies, as in the instance of particle collisions, the 
notion of particle states as such may be lost in any case, 
and recovered only at 
an asymptotic level, i.e., on the tangent space. For such 
generalisations the notion of temperature we have outlined above 
would nonetheless 
survive, indicating a strong interlink between the geometrical 
structure of the state manifold and the thermodynamics 
of associated statistical ensembles. \par 

\subsection{Spin one-half particle} 

In this section, we study the thermal dynamics of 
systems having two energy levels. For such systems, it is easy to see 
that most physical quantities of interest depend only on the energy 
difference but not on the actual values of the energies. Therefore, 
for practical purposes any two level system can be viewed as 
essentially equivalent to a system consisting of a spin one-half 
particle interacting with an external field. \par 

We study the classical situation first. In this case, we have an 
Ising spin in a constant magnetic field whose strength is $h$. 
The associated real Hilbert space ${\cal H}^{2}$ is two 
dimensional, with orthogonal axes given by the spin-up state 
$u^{a}_{\uparrow}$ and the spin-down state $u^{a}_{\downarrow}$. 
If we view ${\cal H}^{2}$ as an $x$-$y$ plane, then the infinite 
temperature ($\beta=0$) state is the intersection of the segment 
$S^{1}_{+}$ of the unit circle with the line $y=x$. As the 
temperature decreases, the equilibrium state moves towards 
$u^{a}_{\uparrow}$ or $u^{a}_{\downarrow}$ depending on the sign 
of the magnetic field $h$. \par 

The unnormalised thermal state $\psi^{a}$ in ${\cal H}^{2}$ can be 
obtained by solving the equation $d\psi^{a}/d\beta = -\frac{1}{2} 
H^{a}_{b}\psi^{b}$, with the result 
\begin{equation} 
\psi^{a}(\beta)\ =\ e^{\beta h} u^{a}_{\uparrow} 
+ e^{-\beta h} u^{a}_{\downarrow}\ . 
\end{equation} 
This can be normalised by dividing the right hand 
side by the partition function $Q(\beta)=2\cosh(\beta h)$. We now 
project this space down to $RP^{1}$, which is effectively a 
circle. Since a circle in ${\cal H}^{2}$ is a double covering of 
$RP^{1}\sim S^{1}$, the two eigenstates are mapped to opposite points 
in $RP^{1}$. \par 

Our geometric `quantisation' procedure for this system is as follows. 
For the given thermal state $\psi^{a}(\beta)$ we assign a phase 
factor, and thus obtain a quantum mechanical state space $CP^{1}$, which 
is viewed as a sphere $S^{2}$. The north and the south 
poles of the sphere correspond to the two energy eigenstates, and 
the great circles passing through these two points correspond to unitary 
equivalent thermal state space trajectories 
parameterised by the phase factor. 
The equator of the sphere, in particular, corresponds to the infinite 
temperature states. This circle is of course a 1-torus, in accordance 
with the general description of the infinite temperature state 
manifold given earlier. The Schr\"odinger dynamics is given by a rigid 
rotation of the sphere about the axis that passes through the 
north and the south poles, the two stationary points. 
This rotation gives rise to 
a Killing vector field on the sphere, where the angular velocity is 
given by the strength of the external field $h$. Thus, the 
temperature value specifies the latitude on the sphere and the 
Schr\"odinger evolution corresponds to a latitudinal circle.  \par 

To pursue this in more detail we introduce a complex 
Hilbert space with coordinates $Z^{\alpha}$ $(\alpha = 0,1)$ which we 
regard as homogeneous coordinates for $CP^{1}$. The complex 
conjugate of $Z^{\alpha}$ is the `plane' ${\bar Z}_{\alpha}$ in 
$CP^{1}$, which in this dimension is simply a point. The point 
corresponding to ${\bar Z}_{\alpha}$ is then given by 
${\bar Z}^{\alpha} = \epsilon^{\alpha\beta}{\bar Z}_{\beta}$, where 
$\epsilon^{\alpha\beta}$ is the natural symplectic form on the 
two-dimensional Hilbert space. The relevant formalism in this case 
is, of course, equivalent to the standard algebra of two-component 
spinors. In particular, by use of the spinor identity $2X^{[\alpha} 
Y^{\beta]} = \epsilon^{\alpha\beta}X_{\gamma}Y^{\gamma}$, where 
$X^{\alpha}\epsilon_{\alpha\beta}=X_{\beta}$, we obtain 
\begin{equation} 
ds^{2}\ =\ \frac{4Z_{\alpha}dZ^{\alpha}
{\bar Z}_{\beta}d{\bar Z}^{\beta}}
{({\bar Z}_{\gamma}Z^{\gamma})^{2}} \label{eq:2df-s} 
\end{equation} 
for the Fubini-Study metric in this situation, and 
\begin{equation} 
Z_{\gamma}dZ^{\gamma}\ =\ i H_{\alpha\beta}Z^{\alpha}
Z^{\beta}dt \label{eq:2dsch} 
\end{equation} 
for the projective Schr\"odinger equation. The Hamiltonian 
$H^{\beta}_{\alpha}$ here has a symmetric representation of the form 
\begin{equation} 
H_{\alpha\beta}\ =\ \frac{2hP_{(\alpha}{\bar P}_{\beta)}}
{({\bar P}_{\gamma}P^{\gamma})}  \label{eq:ham} 
\end{equation} 
where $P^{\alpha}$ and ${\bar P}^{\alpha}$ correspond to the 
stationary points. In particular, we have $H^{\alpha}_{\beta}
P^{\beta}=hP^{\alpha}$ and $H^{\alpha}_{\beta}{\bar P}^{\beta}=
-h{\bar P}^{\alpha}$, which follow if we bear in mind the identity 
${\bar P}_{\gamma}P^{\gamma}=-P_{\gamma}{\bar P}^{\gamma}$. \par 

By insertion of (\ref{eq:ham}) into (\ref{eq:2dsch}) and then 
(\ref{eq:2dsch}) into (\ref{eq:2df-s}), we deduce that 
\begin{equation} 
\left( \frac{ds}{dt}\right)^{2}\ =\ 16h^{2} 
{\rm Prob}[Z^{\alpha}\rightarrow P^{\alpha}] 
{\rm Prob}[Z^{\alpha}\rightarrow{\bar P}^{\alpha}]\ , 
\end{equation} 
where ${\rm Prob}[Z^{\alpha}\rightarrow P^{\alpha}]$ is the 
transition probability from $Z^{\alpha}$ to the north pole 
$P^{\alpha}$, given by the cross ratio 
\begin{equation} 
\frac{(\epsilon_{\alpha\beta}Z^{\alpha}{\bar P}^{\beta})
(\epsilon_{\gamma\delta}{\bar Z}^{\gamma}P^{\delta})}
{(\epsilon_{\alpha\beta}Z^{\alpha}{\bar Z}^{\beta})
(\epsilon_{\gamma\delta}P^{\gamma}{\bar P}^{\delta})}\ =\ 
\frac{1}{2}(1 + \cos\theta)\ .  
\end{equation} 
Here, $\theta$ is the distance from $Z^{\alpha}$ to $P^{\alpha}$, 
given by the usual angular coordinate measured down from the 
north pole. The complementary probability 
${\rm Prob}[Z^{\alpha}\rightarrow{\bar P}^{\alpha}]$ 
is given by $\frac{1}{2}(1 - \cos\theta)$, and it follows that 
the velocity of the trajectory through the state space is 
\begin{equation} 
\frac{ds}{dt}\ =\ 2h\sin\theta\ , 
\end{equation} 
a special case of the Anandan-Aharonov relation \cite{aa} 
noted above. For the evolutionary trajectory we obtain  
\begin{equation} 
Z^{\alpha}\ =\ \cos\frac{\theta}{2} e^{i(ht+\phi)}P^{\alpha} + 
\sin\frac{\theta}{2} e^{-i(ht+\phi)}{\bar P}^{\alpha} \ , 
\end{equation} 
where $\theta$ and $\phi$ are the initial coordinates on the sphere 
for $t=0$. If appropriate, $Z^{\alpha}$ can be normalised by setting 
$P^{\alpha}{\bar P}_{\alpha}=1$. 
A short calculation then shows that the expectation $E$ of the 
energy is given by $E=h\cos\theta$, and that the variance of the 
energy is given by $h^{2}\sin^{2}\theta$. \par 

Now, we are in a position to examine the primitive thermal 
trajectories associated with the spin one-half case. In this case the 
thermal equation is given by 
\begin{equation} 
\frac{dZ^{\alpha}}{d\beta}\ =\ -\frac{1}{2} 
{\tilde H}^{\alpha}_{\beta}Z^{\beta}\ , 
\end{equation} 
where ${\tilde H}^{\alpha}_{\beta} = H^{\alpha}_{\beta} - 
\delta^{\alpha}_{\beta}H^{\mu}_{\nu}Z^{\nu}{\bar Z}_{\mu}
/Z^{\gamma}{\bar Z}_{\gamma}$. By use of (\ref{eq:ham}) we then 
obtain 
\begin{equation} 
Z^{\alpha}\ =\ \frac{e^{-\frac{1}{2}\beta h+i\phi}P^{\alpha} 
+ e^{\frac{1}{2}\beta h-i\phi}{\bar P}^{\alpha}}
{(e^{-\beta h}+e^{\beta h})^{1/2}}\ , 
\end{equation} 
where we assume the normalisation $P^{\alpha}{\bar P}_{\alpha}=1$. 
This shows that at infinite temperature $(\beta=0)$ the state lies 
on the equator, given by $Z^{\alpha}=e^{i\phi}P^{\alpha} + 
e^{-i\phi}{\bar P}^{\alpha}$, where $\phi$ lies in the interval 
from $0$ to $2\pi$. Zero temperature state is obtained by taking the 
limit $\beta\rightarrow\infty$, and we find that $Z^{\alpha}$ 
approaches ${\bar P}^{\alpha}$, the south pole, provided $h>0$. 
For the expectation of the energy $E= H^{\alpha}_{\beta}
Z^{\beta}{\bar Z}_{\alpha}$ we have 
\begin{equation} 
E\ =\ h\frac{e^{-\beta h}-e^{\beta h}}{e^{-\beta h} + 
e^{\beta h}}\ \label{eq:tanh} 
\end{equation} 
which, as expected, ranges from $0$ to $-h$ as $\beta$ ranges from 
$0$ to $\infty$. We note, in particular, that $E$ 
is independent of the phase angle $\phi$, and that the relation 
$E=h\tanh(\beta h)$ agrees with the result for a classical spin. \par 

For other observables this is not necessarily the case, and we have 
to consider averaging over the random state ${\bf Z}^{\alpha}$ 
obtained by replacing $\phi$ with a random variable ${\bf \Phi}$, 
having a uniform distribution over the interval $(0,2\pi)$. Then 
for the density matrix $\rho^{\alpha}_{\beta}:= E[{\bf Z}^{\alpha} 
{\bar {\bf Z}}_{\beta}]$, where $E[-]$ is the unconditional 
expectation, we obtain 
\begin{equation} 
\rho^{\alpha}_{\beta}\ =\ \frac{e^{-\beta h}P^{\alpha}
{\bar P}_{\beta} - e^{\beta h}{\bar P}^{\alpha}P_{\beta}}
{e^{-\beta h} + e^{\beta h}}\ . 
\end{equation} 
The fact that $\rho^{\alpha}_{\beta}$ has trace unity follows from 
the normalisation condition $Z^{\alpha}{\bar Z}_{\alpha}=1$, and the 
identity $Z^{\alpha}{\bar Z}_{\alpha}=-Z_{\alpha}{\bar Z}^{\alpha}$. 
For the energy expectation $\rho(H)=H^{\alpha}_{\beta} 
\rho^{\beta}_{\alpha}$ we then recover (\ref{eq:tanh}). \par 

Alternatively, we can consider the phase-space volume approach 
considered earlier, by assuming a microcanonical distribution for 
this system. Now, the phase space volume of the energy surface 
${\cal E}^{1}_{E}$ (a latitudinal circle) is given by 
${\cal V}(E) = 2\pi\sin\theta$, where $\theta$ is the angle measured 
from the pole of $CP^{1}\sim S^{2}$. Hence, by use of 
(\ref{eq:temp}), along with the energy expectation $E=h\cos\theta$, 
we deduce that the value of the system 
temperature is  
\begin{equation} 
\beta(E)\ =\ \frac{E}{E^{2}-h^{2}}\ . \label{eq:beta}  
\end{equation} 
Since $E\leq0$ and $E^{2}\leq h^{2}$, the inverse temperature $\beta$ 
is positive. Furthermore, we see that $E=0$ implies $\theta=\pi/2$, 
the equator of the sphere, which gives infinite temperature 
($\beta=0$), and $E^{2}=h^{2}$ corresponds to $\theta=\pi$, which 
gives the zero temperature ($\beta=\infty$) state. The equation of 
state for this system can also be obtained by use of the standard 
relation $\beta p=\partial S/\partial {\cal V}$. In this case, we 
obtain the equation of state for an ideal gas, i.e., 
$p{\cal V}=\beta^{-1}$. Explicitly, we have 
\begin{equation} 
p(E)\ =\ -\frac{h}{2\pi E} \sqrt{h^{2}-E^{2}}\ . 
\end{equation} 
The pressure is minimised when the spin aligns with the external 
field, and is maximised at the equator. We note that for positive 
energies $E>0$ the temperature takes negative values. If we take 
$E<0$ and then flip the direction of the external field $h$, 
this situation can be achieved in practice Although the concept of 
such a negative temperature is used frequently in the study of 
Laser phenomena, it is essentially a transient phenomenon 
\cite{kubo}, and is thus not as such an objective of 
thermodynamics. \par 

We note, incidentally, that the distinct energy-temperature 
relationships obtained here 
in equation (\ref{eq:tanh}) for the canonical ensemble and in 
(\ref{eq:beta}) for the microcanonical ensemble, have qualitatively 
similar behaviour. Indeed, for a system consisting of a large number 
of particles these two results are expected to agree in a suitable 
limit. \par 

\section{Discussion} 

The principal results of this paper are the following. First, we 
have formulated a projective geometric characterisation for classical 
probability states. By specialising then to the canonical ensemble 
of statistical mechanics, we have been able to determine the main 
features of thermal trajectories, which are expressed in terms 
of a Hamiltonian gradient flow. This flow is then shown to be a special 
case of a projective automorphism on the state space when it is 
endowed with the natural $RP^{n}$ metric. 
It should be clear that the same formalism, 
and essentially the same results, apply also to the grand canonical 
and the pressure-temperature distributions. \par 

The quantum mechanical dynamics of equilibrium thermal states can be 
studied by consideration of the Hopf-type map $RP^{2n+1} \rightarrow 
CP^{n}$, which in the present context allows one to regard the quantum 
mechanical state space as the base space in a fibre manifold which 
has the structure of an essentially classical thermal state space. 
The 
fact that a projective automorphism on a space of constant curvature 
can be decomposed into two distinct terms suggests the identification 
of the Killing term with the Schr\"odinger evolution of the 
Hamiltonian gradient flow with respect to the symplectic structure, 
and the other term with thermal evolution of the Hamiltonian 
gradient flow with respect to the metric---the former gives rise to 
a linear transformation, while the latter is nonlinear.\par 

There are a number of problems that still remain. First, much of our 
formulation has been based on the consideration of finite dimensional 
examples. The 
study of phase transitions, however, will require a more careful and 
extensive treatment of the infinite dimensional case. Our analysis 
of the projective automorphism group, on the other hand, suggests 
that the infinite dimensional case can also be handled comfortably 
within the geometric framework. Also, for most of the 
paper we have adopted the Schr\"odinger picture, which has perhaps 
the disadvantage of being inappropriate for 
relativistic covariance. It would be desirable to 
reformulate the theory in a covariant manner, in order to study the 
case of relativistic fields.  \par 

Nevertheless, as regards the first problem noted above, the present 
formulation is sufficiently rich in order to allow us to speculate 
on a scenario for the spontaneous symmetry breaking of, say, a pure 
gauge group, in the infinite dimensional situation. In such cases the 
hypersurface of the parameter space (a curve in the one-parameter 
case considered here) proliferates into possibly infinite, 
thermally inequivalent hypersurfaces, corresponding to the 
multiplicity of the ground state degeneracy, at which the symmetry 
is spontaneously broken. The hyper-line characterising the 
proliferation should presumably be called the spinodal line 
(cf. \cite{br}), along which the Riemann curvature of the parameter 
space manifold is expected to diverge. Furthermore, a pure thermal 
state in the `high temperature' region should evolve into a mixed state, 
obtained by averaging over all possible surfaces, by passing 
through the geometrical singularity (the spinodal boundary) 
characterising phase transitions. By a suitable measurement that 
determines which one of the ground states the system is in, this 
mixed state will reduce back to a pure state. In a cosmological 
context this proliferation may correspond, for example, to different 
$\theta$-vacuums \cite{sachs}. It is interesting to note that the 
situation is analogous to the choice of a complex 
structure \cite{gibbons2} for the field theory associated with 
a curved space-time, as the universe evolves. \par 

In any case, the remarkable 
advantages of the use of projective space should 
be stressed. As we have observed, the structure of 
projective space allows us to identify probabilistic operations 
with precise geometric relations. One of the problems involved in 
developing nonlinear (possibly relativistic) generalisations of 
quantum mechanics concerns their 
probabilistic interpretation. Formulated in a 
projective space, such generalisations can be obtained, for 
example, by replacing the Hamiltonian function by a more general 
function, or by introducing a more general metric structure. In this 
way, the assignment of a suitable probability theory can be 
approached in an appropriate way. In particular, as we have 
observed, the canonical ensemble of statistical mechanics has 
an elegant characterisation in projective space---but this is an 
example of a theory that is highly nonlinear and yet purely 
probabilistic. \par 

It is also interesting to observe that the nonlinear generalisation 
of quantum mechanics considered by Kibble \cite{kibble} and 
others can be 
applied to the thermal situation, in the sense that Hamiltonian 
function defined on the state manifold can be replaced by a 
general observable. For such generalisations it is not clear what 
physical interpretation can be assigned. Naively, one might expect 
that by a suitable choice of an observable the resulting 
trajectory characterises some kind of nonequilibrium process. \par 

One of the goals of this paper has been to formulate quantum 
theory at finite temperature in such a way as to allow for the 
possibility of various natural 
generalisations. These might include, for example, the stochastic 
approach to describe measurement theory, or nonlinear relativistic 
extensions of standard quantum theory, as noted above. These 
generalisations will be pursued further elsewhere. \par 

\vspace{0.4cm} 

\noindent {\bf Acknowledgement.} DCB is grateful to Particle 
Physics and Astronomy Research Council for financial support. \par 

\vspace{0.4cm} 
$*$ Electronic address: d.brody@damtp.cam.ac.uk \par 
$\dagger$ Electronic address: lane@ml.com\par

\end{document}